\documentclass[12pt]{article}
\usepackage[margin=1in]{geometry}
\usepackage[utf8]{inputenc}

\usepackage{amsmath,amssymb,amsfonts,amsthm}

\usepackage{breqn}

\usepackage{bold-extra}

\usepackage{graphicx}
\usepackage{xcolor}
\usepackage{url}
\usepackage{hyperref}
\hypersetup{
  colorlinks   = true, 
  urlcolor     = blue, 
  linkcolor    = black, 
  citecolor   = black 
}
\usepackage[numbers,sort&compress]{natbib}
\let\OLDthebibliography\thebibliography
\renewcommand\thebibliography[1]{
  \OLDthebibliography{#1}
  \setlength{\parskip}{4pt}
  \setlength{\itemsep}{0pt plus 0.3ex}
}

\allowdisplaybreaks
\usepackage{tabularx}
\usepackage{bbm}
\usepackage{listings}
\usepackage{longtable}
\usepackage{array}

\usepackage{multicol}
\usepackage{enumitem}

\newcommand\refeq[1]{Eq.~(\ref{#1})}
\newcommand\refeqs[1]{Eqs.~(\ref{#1})}
\newcommand\refta[1]{Tab.~\ref{#1}}
\newcommand\refse[1]{Sect.~\ref{#1}}

\newcommand\citere[1]{Ref.~\cite{#1}}
\newcommand\citeres[1]{Refs.~\cite{#1}}
\newcommand\refap[1]{App.~\ref{#1}}
\def\reffi#1{\mbox{Fig.~\ref{#1}}}

\usepackage{soul}

\newcommand{\gev}{\ \mathrm{GeV}}
\newcommand{\tev}{\ \mathrm{TeV}}

\newcommand{\ii}{\text{i}}

\begin{document}

\thispagestyle{empty}

\def\thefootnote{\fnsymbol{footnote}}

\begin{flushright}
  DESY 21-228 ~~ \\
  IFT-UAM-CSIC-153 ~~
\end{flushright}

\vspace*{1cm}

\begin{center}
  {\Large Vacuum (meta-)stability
    in the $\mu\nu$SSM}

  \vspace{1cm}

  Thomas Biekötter\footnote{thomas.biekoetter@desy.de}\\[0.2em]
  {\textit{
   Deutsches Elektronen-Synchrotron DESY, Notkestra{\ss}e 85, 22607 Hamburg, Germany\\[0.8em]
 }}
 Sven Heinemeyer\footnote{Sven.Heinemeyer@cern.ch}\\[0.2em]
 {\textit{
   Instituto de Física Teórica UAM-CSIC, Cantoblanco, 28049,
   Madrid, Spain\\[0.8em]
 }}
 Georg Weiglein\footnote{georg.weiglein@desy.de}\\[0.2em]
 {\textit{
   Deutsches Elektronen-Synchrotron DESY, Notkestra{\ss}e 85, 22607 Hamburg, Germany\\[0.2em]
   II. Institut für Theoretische Physik, Universität Hamburg,
    Luruper Chaussee 149, 22607 Hamburg, Germany\\[0.4em]
  }}

\vspace*{1cm}

\begin{abstract}
We perform an analysis of the vacuum stability
of the neutral scalar potential of
the \mbox{$\mu$-from-$\nu$} Supersymmetric Standard Model
($\mu\nu$SSM).
As an example scenario, we
discuss the alignment without
decoupling limit of the $\mu\nu$SSM,
for which the required conditions on
the Higgs sector are derived.
We demonstrate that in this limit large parts
of the parameter space
feature
unphysical minima that are deeper than the electroweak
minimum.
In order to estimate the lifetime of
the electroweak vacuum, we calculate
the rates for the
tunneling process into each unphysical
minimum.
We find that 
in many cases the resulting lifetime is longer than
the age of the universe,
such that the considered parameter region 
is not excluded.
On the other hand, we also find
parameter regions in which the
electroweak vacuum is short-lived. We
investigate how the different regions 
of stability
are related to the presence of
light right-handed sneutrinos.
Accordingly, a vacuum stability analysis that 
accurately takes into account the possibility of long-lived metastable 
vacua is crucial for a reliable assessment of the phenomenological 
viability of the parameter space
of the $\mu\nu$SSM and its resulting
phenomenology at the (HL)-LHC.
\end{abstract}

\end{center}

\renewcommand{\thefootnote}{\arabic{footnote}}
\setcounter{footnote}{0} 

\newpage

\section{Introduction}
\label{intro}
The Standard Model (SM) of particle
physics successfully describes the
fundamental interactions of the known
matter fields 
at the present level of experimental accuracy.
In particular, the discovery of a scalar
particle with a mass of $\approx 125\gev$
at the Large Hadron Collider
(LHC)~\cite{Aad:2012tfa,Chatrchyan:2012ufa},
that within the current
experimental
uncertainties~\cite{Khachatryan:2016vau,
Aad:2019mbh,Sirunyan:2018koj}
behaves as
predicted by the realisation of the Brout-Englert-Higgs
mechanism within
the SM~\cite{Englert:1964et,Higgs:1964pj},
has confirmed the existence of a non trivial
vacuum structure of the universe.
Despite the great success of the SM,
there are also phenomena that cannot be
accommodated, and furthermore the
SM has conceptual shortcomings.
In particular, the SM does not provide a
sizable contribution to the relic
abundance of cold dark matter,
the neutrino oscillation data cannot
be explained with strictly massless
SM neutrinos,
there is no mechanism to 
account for the observed
matter-antimatter asymmetry
of the universe,
and there is no consistent formulation
of the SM that includes gravitational
interactions.

Another conceptual problem of the SM is related
to the fact that its scalar sector,
incorporating a single SU(2) doublet scalar field,
just provides a minimal parameterization of electroweak symmetry breaking. 
The renormalizable scalar
potential of the SM can be expressed in terms of only
two parameters, e.g., the vacuum expectation value (vev),
$v \approx 246\gev$, and 
the mass of the predicted Higgs boson, which within the SM is identified with the mass of the 
detected particle, $m_{h_{125}} \approx 125\gev$.
This minimal construction is sufficient
to generate the masses of the
gauge bosons, the Higgs boson, the
quarks and the charged leptons
via spontaneous symmetry breaking.
However, within the SM there is no dynamical
origin of the Higgs potential, and
the mass of the Higgs boson of the SM
is not protected by symmetries against
large corrections from physics at high scales.
This gives rise to the
so-called \textit{hierarchy
problem}~\cite{Weinberg:1975gm,Gildener:1976ai}
regarding the observed value
of $m_{h_{125}}$ and more generally
the stability of the electroweak (EW) scale.

One of the most thoroughly studied
frameworks of BSM physics is Supersymmetry (SUSY).
Models based on SUSY provide a solution for
the hierarchy problem, 
since the symmetry between
bosonic and fermionic degrees
of freedom leads to a cancellation between 
contributions with
a quadratic dependence on new
physics at higher
energies~\cite{Veltman:1980mj,
Dimopoulos:1981zb,Witten:1981nf},
thus protecting the masses of the
Higgs bosons of the model.
Furthermore, for the observed gauge
bosons and the states in the 
Higgs sector SUSY predicts new fermionic superpartners, 
as well as new bosonic superpartners for
the three generations of quarks and leptons.
As a result, SUSY extensions of the SM have
a substantially richer matter sector,
and in addition to the hierarchy
problem many other shortcomings
of the SM can be addressed as well. For instance,
there are several 
possible
candidates for cold dark matter.

A particularly well motivated
SUSY model is the $\mu$-from-$\nu$
Supersymmetric Standard Model
($\mu\nu$SSM)~\cite{Bratchikov:2005vp,
Munoz:2009an}.\footnote{See also
\citere{Lopez-Fogliani:2020gzo} for a
recent review of the $\mu\nu$SSM.}
Beyond the well-known appealing
features of commonly studied
SUSY models, in the $\mu\nu$SSM
the tiny neutrino
masses and their mixings can
be accommodated via an EW
seesaw mechanism, where it is
required that the matter content is
enlarged w.r.t.\ the SM
by right-handed neutrinos~\cite{Escudero:2008jg,Ghosh:2008yh,
Fidalgo:2009dm,Bartl:2009an,Ghosh:2010zi}.
Their superpartners,
the ``right-handed''\footnote{This
phrase is used also for the scalar particles 
in order to indicate 
that they are the superpartners of right-handed fermions.}
scalar neutrinos (sneutrinos), 
are gauge singlet scalar fields.
If the right-handed sneutrinos acquire
vevs, the so-called $\mu$-term of the
Minimal Supersymmetric Standard
Model (MSSM) can be generated
effectively, in
complete analogy to the $Z_3$
symmetric Next-to-MSSM (NMSSM).
Consequently, the $\mu\nu$SSM also provides
a solution to the so-called
$\mu$-problem~\cite{Ellis:1988er,Miller:2003ay}.
By construction, the $\mu\nu$SSM
does not allow for a consistent
assignment of conserved $R$ parity
charges, giving rise to the fact
that there is no stable
lightest SUSY particle.
Compared to the (N)MSSM, the collider
constraints from the LHC are
therefore substantially
weaker~\cite{Ghosh:2014ida,Ghosh:2017yeh,
Lara:2018rwv,Lara:2018zvf,Kpatcha:2019qsz,
Kpatcha:2019gmq}.
Regarding dark matter,
it is possible to accommodate
the measured relic abundance
by means of decaying, but long-lived,
gravitinos or axinos~\cite{Gomez-Vargas:2016ocf,
Gomez-Vargas:2019vci,
Gomez-Vargas:2019mqk}.

One of the key features of the
$\mu\nu$SSM is the mixing of the
two Higgs doublet fields $H_d$ and
$H_u$ with the three left-handed and
right-handed sneutrino fields
$\widetilde{\nu}_{iL}$ and
$\widetilde{\nu}_{iR}$, leading
to a total of eight neutral 
CP-even
particles and seven 
neutral CP-odd scalar
particles.\footnote{We assume CP conservation
throughout this paper.}
While the mixing of the left-handed
sneutrinos with the Higgs doublet fields
is suppressed by the smallness of lepton
number breaking, the mixing of the right-handed sneutrinos 
with the Higgs fields can
be sizable, while being in agreement with the current
experimental limits~\cite{Kpatcha:2019qsz,
Biekotter:2019gtq}.
The phenomenological
consequences of the latter possibility have been
thoroughly studied, paying
special attention to the prediction
for the mass of the SM-like
Higgs boson~\cite{Biekotter:2017xmf,Biekotter:2019gtq,
Biekotter:2020ehh}. In addition to
the more complicated scalar
particle spectrum, the enlarged
Higgs sector compared to the SM and
the (N)MSSM also results in a substantially
more complicated scalar potential.
While it is always possible to ensure
the existence of a phenomenologically
viable EW vacuum by conveniently choosing
the vevs of the neutral scalar fields
as input parameters (as will be discussed 
in \refse{intromunu}), the scalar potential of
the $\mu\nu$SSM in general contains further
unphysical local minima. If one or more
of the unphysical minima are deeper than
the EW minimum, the question arises
whether the EW vacuum can become
short-lived in comparison to the age
of the universe, and if this is the case
the corresponding
parameter point should be rejected.

Up to now no detailed analysis of constraints
on the parameter space of the $\mu\nu$SSM
arising from possible instabilities of the
EW vacuum has been carried out in the literature.
To the best of our knowledge,
only \citere{Escudero:2008jg}
took into account such constraints,
demanding, however, that the EW minimum
is the global minimum of the potential, 
i.e.\ the possibility of a long-lived
metastable EW vacuum was not taken
into account.
Apart from that, heuristic constraints
that have been obtained for
the presence of color-breaking unphysical 
minima in the MSSM could be
considered~\cite{Casas:1995pd}.
Besides their limited reliability,
constraints of this kind 
do not capture the possibility that
charge- and color-conserving
minima can nevertheless be unphysical.
Accordingly, we study in this
paper the impact of
genuine effects of the $\mu\nu$SSM on
the stability of the EW vacuum.

In this paper, we 
perform a detailed
analysis of the neutral scalar potential
of the $\mu\nu$SSM regarding
EW vacuum instabilities.
In a first step, we determine
the unphysical minima of the tree-level
neutral scalar potential by making use of an
homotopy continuation method.
In a second step,
we estimate the rates for the decays of the EW
vacuum into all unphysical minima
below the EW minimum using a semi-classical
approximation. By subsequently comparing these
rates to the age of the universe, we
categorize parameter points into ones
with a short-lived
(unstable), long-lived (metastable)
and an absolutely stable EW vacuum, 
and we discuss the impact of the resulting
constraints on
the parameter space of the model.
As an example scenario, 
in our numerical analysis
we apply
the investigation of vacuum stability constraints to the
\textit{alignment-without-decoupling
limit} of the $\mu\nu$SSM. It turns out that
in this limit the
EW minimum is not the global minimum
of the potential. However, in large parts
of the analyzed parameter regions the
decay rates into the unphysical minima
are so small that the EW vacuum is long-lived 
compared to the age of the universe,
and those parameter regions with
a metastable vacuum are 
therefore phenomenologically viable.

The paper is organized as follows:
In \refse{intromunu} we introduce the Higgs sector
of the $\mu\nu$SSM, followed by a discussion
of the alignment without decoupling limit
in \refse{alignwodec}.
In \refse{vacstab} we describe the determination of
the minima of the neutral scalar
potential and the procedure that is applied for
assessing the stability of the EW vacuum.
In \refse{numana} we apply this approach to
example scenarios in the alignment limit.
Finally, we summarize our results and give
a brief outlook on possible avenues for
future research in \refse{conclu}. 
In the appendix we briefly describe the public code 
\texttt{munuSSM} that can be used for performing
the vacuum stability test in the
$\mu\nu$SSM parameter space.
We furthermore provide explicit expressions for the 
coefficients of the scalar potential relative
to the EW vacuum for 
some special cases.

\section{The neutral scalar sector of the 
\texorpdfstring{\boldmath{$\mu\nu$}SSM}{munuSSM}}
\label{intromunu}
The neutral scalar
potential of the $\mu\nu$SSM 
consists of contributions from
the superpotential $W$ and the soft SUSY-breaking
Lagrangian $\mathcal{L}_{\rm soft}$.
The part of $W$ that contains
the neutral superfields is given
by~\cite{Bratchikov:2005vp,Munoz:2009an}
\begin{align}
W = \; &
\epsilon_{ab} \left(
Y^{\nu}_{ij} \, \hat H_u^b\, \hat L^a_i \, \hat \nu^c_j 
-
\lambda_{i} \, \hat \nu^c_i\, \hat H_u^b \hat H_d^a
\right)
+
\frac{1}{3}
\kappa{_{ijk}} 
\hat \nu^c_i\hat \nu^c_j\hat \nu^c_k + \dots
\ ,
\label{superpotential}
\end{align}
where $\hat H_d^T=(\hat H_d^0, \hat H_d^-)$ and 
$\hat H_u^T=(\hat H_u^+, \hat H_u^0)$ are the Higgs doublet
superfields,
$\hat L_i$
are the left-chiral
lepton superfields with the left-chiral
neutrino superfields $\hat\nu_i$ in the
upper component,
and $\hat{\nu}^c$ are
the right-chiral neutrino superfields.
${i,j=1,2,3}$ are the family indices, and
${a,b=1,2}$ are the weak isospin indices
$(\epsilon_{12}=1)$.
The portal couplings $\lambda_i$
give rise to the mixing between the right-handed
sneutrinos and the Higgs doublet fields,
and also the $\mu$-term
is generated after electroweak symmetry
breaking (EWSB)
($\mu=\lambda_i v_{iR} / \sqrt{2}$).
The self-couplings $\kappa_{ijk}$
generate masses for the right-handed
(s)neutrinos after EWSB, such that they
determine the scale of the seesaw mechanism.
Since the seesaw scale is in this way
related to the EW scale,
for the neutrino Yukawa couplings
one has to demand
$Y^\nu_{ij} \lesssim Y_e \approx 10^{-6}$,
with $Y_e$ the electron Yukawa coupling,
in order to obtain left-handed neutrino
masses at the sub-eV level.

The part of $\mathcal{L}_{\rm soft}$
that contains the neutral scalar fields
can be written
as~\cite{Bratchikov:2005vp,Munoz:2009an}
\begin{align}
-\mathcal{L}_{\text{soft}}  & = 
\epsilon_{ab} \left(
T^{\nu}_{ij} \, H_u^b \, \widetilde L^a_{iL} \widetilde \nu_{jR}^*
-
T^{\lambda}_{i} \, \widetilde \nu_{iR}^*
\, H_d^a  H_u^b
+
\frac{1}{3} T^{\kappa}_{ijk} \, \widetilde \nu_{iR}^*
\widetilde \nu_{jR}^*
\widetilde \nu_{kR}^*
\
+ \text{h.c.}\right)
\nonumber\\
&+ \
\left(m_{\widetilde{L}}^2\right)_{ij}  
\widetilde{L}_{iL}^{a*}  
\widetilde{L}^a_{jL}
+
\left(m_{\widetilde{\nu}}^2\right)_{ij} \widetilde{\nu}_{iR}^*
\widetilde\nu_{jR} 
+ 
m_{H_d}^2 {H^a_d}^*
H^a_d + m_{H_u}^2 {H^a_u}^*
H^a_u
+ \dots \ ,
\label{Vsoft}
\end{align}
where $H_d$, $H_u$,
$\widetilde{\nu}_{iL}$ and
$\widetilde{\nu}_{iR}$ denote the
scalar components of the superfields
$\hat H_d$, $\hat H_u$,
$\hat \nu_i$ and $\hat \nu_i^c$,
respectively.
As usual, soft mass parameters that are non-diagonal
in field space are neglected.
Thus, terms of the form
$( (m_{H_d\widetilde{L}}^2)_i
H_d^{a*} \widetilde{L}_{iL}^a +
\mathrm{h.c.} )$ are
absent, and the matrices $m_{\widetilde{L}}^2$
and $m_{\widetilde{\nu}}^2$
are assumed to be diagonal.
For the trilinear scalar couplings,
we will make use of the decomposition
$T^\nu_{ij}=A^\nu_{ij} Y^\nu_{ij}$,
$T^\lambda_i = A^\lambda_i \lambda_i$,
$T^\kappa_{ijk} = A^\kappa_{ijk} \kappa_{ijk}$,
which is motivated in models of
supergravity with diagonal K\"ahler
metric~\cite{Brignole:1998dxa}.

The soft terms shown above, in combination
with the $D$- and $F$-terms derived
from the superpotential, give rise to
the neutral scalar potential
\begin{align}
V = V_{\text{soft}} + V_F  &+  V_D\ , \;
\text{with}
\label{eq:fullpot} \\[0.2em]
V_{\text{soft}}= \; & 
\left(
T^{\nu}_{ij} \, H_u^0\,  \widetilde \nu_{iL} \, \widetilde \nu_{jR}^* 
- T^{\lambda}_{i} \, \widetilde \nu_{iR}^*\, H_d^0  H_u^0
+ \frac{1}{3} T^{\kappa}_{ijk} \, \widetilde \nu_{iR}^* \widetilde \nu_{jR}^* 
\widetilde \nu_{kR}^*\
+
\text{h.c.} \right)
\nonumber\\
+ \; &
\left(m_{\widetilde{L}}^2\right)_{ij} \widetilde{\nu}_{iL}^* \widetilde\nu_{jL}
+
\left(m_{\widetilde{\nu}}^2\right)_{ij} \widetilde{\nu}_{iR}^* \widetilde\nu_{jR} +
m_{H_d}^2 {H^0_d}^* H^0_d + m_{H_u}^2 {H^0_u}^* H^0_u
\ ,
\label{akappa}
\\[0.2em]
V_{F}  = \; &
 \lambda_{j}\lambda_{j} H^0_{d}H_d^0{^{^*}}H^0_{u}H_u^0{^{^*}}
 +
\lambda_{i}\lambda_{j} \tilde{\nu}^{*}_{iR}\tilde{\nu}_{jR}H^0_{d}H_d^0{^*}
 +
\lambda_{i}\lambda_{j}
\tilde{\nu}^{*}_{iR}\tilde{\nu}_{jR}  H^0_{u}H_u^0{^*}   
\nonumber\\
+ \; &
\kappa_{ijk}\kappa_{ljm}\tilde{\nu}^*_{iR}\tilde{\nu}_{lR}
                                   \tilde{\nu}^*_{kR}\tilde{\nu}_{mR}
- \left(\kappa_{ijk}\lambda_{j}\tilde{\nu}^{*}_{iR}\tilde{\nu}^{*}_{kR} H_d^{0*}H_u^{0*}
 -Y^{\nu}_{ij}\kappa_{ljk}\tilde{\nu}_{iL}\tilde{\nu}_{lR}\tilde{\nu}_{kR}H^0_{u}
 \right.
 \nonumber\\
 + \; &
 \left.
 Y^{\nu}_{ij}\lambda_{j}\tilde{\nu}_{iL} H_d^{0*}H_{u}^{0*}H^0_{u}
+{Y^{\nu}_{ij}}\lambda_{k} \tilde{\nu}_{iL}^{*}\tilde{\nu}_{jR}\tilde{\nu}_{kR}^* H^0_{d}
 + \text{h.c.}\right) 
\nonumber \\
+ \; &
Y^{\nu}_{ij}{Y^{\nu}_{ik}} \tilde{\nu}^{*}_{jR}
\tilde{\nu}_{kR}H^0_{u}H_u^0{^*}
 +
Y^{\nu}_{ij}{Y^{\nu}_{lk}}\tilde{\nu}_{iL}\tilde{\nu}_{lL}^{*}\tilde{\nu}_{jR}^{*}
                                  \tilde{\nu}_{kR}  
 +
Y^{\nu}_{ji}{Y^{\nu}_{ki}}\tilde{\nu}_{jL}\tilde{\nu}_{kL}^* H^0_{u}H_u^{0*}\, ,
\\[0.2em]
V_D  = \; &
\frac{1}{8}\left(g_1^{2}+g_2^{2}\right)\left(\widetilde\nu_{iL}\widetilde{\nu}_{iL}^* 
+H^0_d {H^0_d}^* - H^0_u {H^0_u}^* \right)^{2}\, .
\label{dterms}
\end{align}
The required existence
of a physical
EW minimum 
can be made explicit by using the decomposition
\begin{equation}
H_{d,u}^0 = \frac{1}{\sqrt 2} \left(H_{d,u}^{\mathcal{R}} + v_{d,u} +
\ii\, H_{d,u}^{\mathcal{I}}\right)\ , \quad
\widetilde{\nu}_{iR,L} =
      \frac{1}{\sqrt 2}
      \left(\widetilde{\nu}^{\mathcal{R}}_{iR,L}+ v_{iR,L} +
        \ii\, \widetilde{\nu}^{\mathcal{I}}_{iR,L}\right) \ ,
\label{vevcompo}
\end{equation}
where the superscripts $^{\mathcal{R}}$ and
$^{\mathcal{I}}$ denote CP--even and CP--odd
components of each scalar field,
respectively.
The numerical prefactor ensures
that the  kinetic terms of the real
fields $H_{d,u}^{\mathcal{R}}$,
$H_{d,u}^{\mathcal{I}}$,
$\widetilde{\nu}^{\mathcal{R}}_{iR,L}$ and
$\widetilde{\nu}^{\mathcal{I}}_{iR,L}$ are
canonically normalized, provided that
the complex scalar fields are canonically
normalized (see the discussion
in \citere{Hollik:2018wrr}).
The EW minimum is then defined by the vevs
\begin{align}
\langle H_d^0 \rangle = \frac{v_d}{\sqrt 2}\ , \, \quad 
\langle H_u^0 \rangle = \frac{v_u}{\sqrt 2}\ , \,
\quad 
\langle \widetilde \nu_{iR}\rangle =
\frac{v_{iR}}{\sqrt 2}\ , \,  \quad
\langle \widetilde \nu_{iL} \rangle =
\frac{v_{iL}}{\sqrt 2} \; .
\label{vevs}
\end{align}
In order to simplify the notation,
the CP-even and the CP-odd
components of the scalar fields
will be collectively denoted by
$\phi = (H_{d}^{\mathcal{R}},
H_{u}^{\mathcal{R}},
\widetilde{\nu}^{\mathcal{R}}_{iR},
\widetilde{\nu}^{\mathcal{R}}_{iL})$
and $\sigma = (H_{d}^{\mathcal{I}},
H_{u}^{\mathcal{I}},
\widetilde{\nu}^{\mathcal{I}}_{iR},
\widetilde{\nu}^{\mathcal{I}}_{iL})$, respectively.
For practical reasons,
we focus only on the CP-even
part of the scalar potential to reduce
the number of field dimensions, 
i.e.\ in our analysis we assume that
$\langle \sigma_i \rangle = 0$,
and the parameters
$v_d,v_u,v_{iR},v_{iL}$ are assumed
to be real.\footnote{Thus, our analysis does not
take into account
the possibility of further CP- or
charge-breaking minima.}

The vevs defined in \refeq{vevcompo}
will be used as input
parameters.\footnote{Regardless
of the prefactors $1 / \sqrt{2}$ in
\refeqs{vevs}, we will refer to
the parameters $v_u$, $v_d$, $v_{iL}$ and $v_{iR}$ as
vevs.}
The soft mass parameters
$m_{H_d}^2$, $m_{H_u}^2$,
$(m_{\widetilde{L}}^2)_{ii}$ and
$(m_{\widetilde{\nu}}^2)_{ii}$
are then given as a function of the vevs demanding that
the minimization conditions
\begin{align}
0=&-m_{H_d}^2v_d -
\frac{1}{8}\left(g_1^2+g_2^2\right)v_d \left( v_d^2 + v_{iL}v_{iL} - v_u^2\right)
-\frac{1}{2} v_d v_u^2 \lambda_i \lambda_i +
\frac{1}{\sqrt{2}} v_u v_{iR} T^\lambda_i
\notag \\
&+\frac{1}{2} v_u^2 Y^\nu_{ji} \lambda_i v_{jL} -
\frac{1}{2} v_d v_{iR} \lambda_i v_{jR} \lambda_j
+\frac{1}{2} v_u \kappa_{ikj} \lambda_i v_{jR} v_{kR} +
\frac{1}{2} v_{iR} \lambda_i v_{jL} Y^\nu_{jk} v_{kR} \; , \label{eq:tp1}\\[4pt]
0=&-m_{H_u}^2v_u+
\frac{1}{8}\left( g_1^2+g_2^2\right)v_u\left( v_d^2+v_{iL}v_{iL}-v_u^2\right)
-\frac{1}{2} v_d^2 v_u \lambda_i \lambda_i +
\frac{1}{\sqrt{2}} v_d v_{iR} T^\lambda_i
\notag \\
&+v_d v_u Y^\nu_{ji} \lambda_i v_{jL} -
\frac{1}{\sqrt{2}} v_{iL} T^\nu_{ij} v_{jR} -
\frac{1}{2} v_u v_{iR} \lambda_i v_{jR} \lambda_j
-\frac{1}{2} v_u Y^\nu_{ji} Y^\nu_{ki} v_{jL} v_{kL}
\notag \\
&-\frac{1}{2} v_u Y^\nu_{ij} Y^\nu_{ik} v_{jR} v_{kR} +
\frac{1}{2} v_d \kappa_{ijk} \lambda_i v_{jR} v_{kR} -
\frac{1}{2} Y^\nu_{li} \kappa_{ikj} v_{jR} v_{kR} v_{lL} \ , \label{eq:tp2}\\[4pt]
0=
&-\left(m_{\widetilde{\nu}}^2\right)_{ij} v_{jR} -
\frac{1}{\sqrt{2}} v_u v_{jL} T^\nu_{ji} -
\frac{1}{2} v_u^2 Y^\nu_{ji} Y^\nu_{jk} v_{kR} +
v_d v_u \kappa_{ijk} \lambda_j v_{kR} -
\frac{1}{\sqrt{2}} T^\kappa_{ijk} v_{jR} v_{kR} \notag \\
&+\frac{1}{2} v_d v_{jL} Y^\nu_{ji} v_{kR} \lambda_k -
v_u Y^\nu_{lj} \kappa_{ijk} v_{kR} v_{lL} -
\frac{1}{2} v_{jL} Y^\nu_{ji} v_{kL} Y^\nu_{kl} v_{lR} -
\kappa_{ijm} \kappa_{jlk} v_{kR} v_{lR} v_{mR} \notag \\
&-\frac{1}{2} \left( v_d^2 + v_u^2 \right) \lambda_i \lambda_j v_{jR} +
\frac{1}{2} v_d v_{jL} Y^\nu_{jk} v_{kR} \lambda_i +
\frac{1}{\sqrt{2}} v_d v_u T^\lambda_i \ , \label{eq:tp3}\\[4pt]
0=&-
\left(m_{\widetilde{L}}^2\right)_{ij}v_{jL} -
\frac{1}{8}\left( g_1^2+g_2^2\right)v_{iL}\left( v_d^2+v_{jL}v_{jL}-v_u^2\right)
+\frac{1}{2} v_d v_u^2 Y^\nu_{ij} \lambda_{j} -
\frac{1}{\sqrt{2}} v_u v_{jR} T^\nu_{ij}
\notag \\
&-\frac{1}{2} v_u^2 Y^\nu_{ij} Y^\nu_{kj} v_{kL} +
\frac{1}{2} v_d v_{jR} Y^\nu_{ij} v_{kR} \lambda_k
-\frac{1}{2} v_u Y^\nu_{ij} \kappa_{jkl} v_{kR} v_{lR} -
\frac{1}{2} v_{jR} Y^\nu_{ij} v_{kL} Y^\nu_{kl} v_{lR} \label{eq:tp4} \ ,
\end{align}
are fulfilled.
Moreover, for phenomenological reasons the
Hessian matrix of $V$ as a function of 
$\phi_i$ and $\sigma_i$ has to be positive definite
in the EW vacuum
in order to avoid tachyonic states. 
It should be noted that
in 
principle a parameter
point with a tachyonic
state at tree-level could 
still be physical, since 
sufficiently large radiative corrections to the scalar
masses might give rise to a change of the
curvature in a particular field
direction, thus leading to the
presence of only non-tachyonic physical
states.
In our analysis, we do not take into
account radiative corrections to the
scalar potential, 
and accordingly we discard
parameter points with tachyonic states at
tree-level.
In general, the inclusion of loop corrections for
the analysis of the vacuum stability
would demand a conceptually different
approach than the one pursued here,
because we make use of the
fact that the tree-level potential
only contains polynomial contributions
(see also \refse{vacstab}).
Here it should be kept in mind that 
the question whether the usual approach of
taking into account higher-order
corrections in the form of an effective
\textit{Coleman-Weinberg}
potential actually improves the calculation
of the lifetime of the EW minimum
is still an open issue. This is related
to the fact that
the effective potential formulation does not capture
momentum-dependent contributions, which
were demonstrated to be
relevant for the calculation of
the decay rates, leading to an
inconsistent perturbative expansion
when truncated at a certain order
of $\hbar$~\cite{Andreassen:2016cff,
Andreassen:2016cvx} (see also
\citere{Hollik:2018wrr} for a discussion).

In addition to the absence of tachyonic states,
it is required for
a phenomenologically viable parameter point
that the
vevs of the fields charged under the EW symmetry
have to fulfill the condition
${v^2 = v_u^2 + v_d^2 + v_{iL} v_{iL} \approx 246\gev}$,
such that the observed values of
the gauge boson masses are 
properly reproduced.
It is convenient to define $\tan\beta = v_u / v_d$ in order 
to make a connection to the MSSM. Using the above
relations, $v$ and $\tan\beta$ can be used as input
parameters instead of $v_d$ and $v_u$.
From \refeq{eq:tp4} one can deduce, using
$T^\nu_{ij} = A^\nu_{ij} Y^\nu_{ij}$, that a
solution to this equation requires that
$v_{iL} \approx Y^\nu_{ii} v$, such that
$v_{iL} \ll v_d,v_u$. The scale of
$v_{iR}$, on the other hand, is given by
the SUSY-breaking scale via the parameters
$(m_{\widetilde{\nu}}^2)_{ij}$, $T^\kappa_{ijk}$
and $T^\lambda_i$ in \refeq{eq:tp3}.

The CP-even part of the scalar potential
$V$ at lowest order is a
polynomial in eight (field) dimensions
with degree four. Consequently, it can have
numerous coexisting local minima besides the
EW minimum.\footnote{The stationary conditions form
a system of eight polynomial equations of degree three
in eight coordinates. Such a system has a total number
of up to $3^8 = 6561$
distinct solutions in the complex plane.
However, the number of
real solutions belonging to a local minimum of the
potential was substantially 
smaller in all cases
considered in our numerical analysis.}
Some of these minima are physically redundant,
because the potential contains accidental
discrete symmetries
yielding degenerate stationary points.
For instance, one can see that if
$\phi = (v_d,v_u,v_{iR},v_{iL})$ is
a solution to \refeqs{eq:tp1}--(\ref{eq:tp4}), then
there is always a second physically
identical solution
at $\phi = (-v_d,-v_u,v_{iR},-v_{iL})$.
However, there can also be solutions of
the minimization conditions that are potentially
\textit{dangerous} for the stability
of the EW vacuum. Any field
configuration $\phi_i$ that solves
\refeqs{eq:tp1}--(\ref{eq:tp4}), that
has a positive definite Hessian matrix,
and for which the value of the potential
in the minimum
is smaller than the one of the EW minimum,
constitutes a
deeper local minimum of the potential.
The existence of such a deeper local minimum gives
rise to the possibility of a rapid EW
vacuum decay.
We will refer to such minima as
\textit{unphysical minima}, since they do
not fulfill $v \approx 246\gev$.
Furthermore, due to the fact that the values
of $v_{iL}$ are different in the unphysical
minima, the predicted neutrino masses are
also modified compared to the
prediction based on the EW minimum.

Consequently, for a phenomenologically viable
parameter point one has to verify that
the EW minimum is either the global minimum
of the potential (but there are constraints that apply
even in this case, see below),
or that, if deeper unphysical minima
exist, the lifetime of the (metastable) EW vacuum is
large in comparison to the age of the universe. As
discussed in \citeres{Baum:2020vfl,Biekotter:2021ysx},
taking into account the thermal
history of models with multiple scalar fields,
one finds that a parameter point featuring 
a global EW minimum at zero temperature
may still be unphysical. This happens if the 
transition to the EW vacuum would have
occurred via a first-order
phase transition, but the transition probabilities
turn out to be
never large enough to allow the
onset of the bubble nucleation
of true EW vacuum bubbles in
the early universe.
In this case, the universe would be
trapped in one of the unphysical vacua
in the limit of zero temperature despite the fact that the 
EW minimum is deeper than the unphysical one.
In our analysis in this paper we restrict ourselves to the case of zero temperature. 
One of the important findings of
our analysis will be that
the rich vacuum structure of the $\mu\nu$SSM neutral scalar
potential gives rise to the fact that several 
unphysical vacua can be present simultaneously, such that 
there are several
possibilities for \textit{vacuum trapping} of 
the universe in 
an unphysical field configuration.
The analysis of the stability of the EW vacuum at $T = 0$,
which is the focus of the present paper, is
carried out under the assumption
that the thermal history of the 
universe has been such that the universe
is actually in the EW minimum 
at $T = 0$. This analysis that we will perform 
can be used to determine constraints on
the parameter space of 
the $\mu\nu$SSM essentially independently
of the question which additional 
constraints would arise from a
dedicated analysis of the thermal history 
of the universe. We leave the latter
kind of analysis, which in particular
takes into account constraints arising
from the above-mentioned
effect of vacuum trapping,
for future work.

Since the multi-dimensional parameter space of the 
$\mu\nu$SSM cannot be covered entirely, we focus our
numerical discussion on the alignment without
decoupling limit, which is theoretically and
experimentally well motivated in view of
the fact that the properties 
of the Higgs boson at $125\gev$ are SM-like.
The conditions that apply 
for the parameter space of the model in this limit and the 
resulting phenomenological features
will be introduced in
the following section.

\section{Alignment without decoupling
in the \texorpdfstring{\boldmath{$\mu\nu$}SSM}{munuSSM}}
\label{alignwodec}
We start by investigating
the alignment without decoupling
conditions for the $\mu\nu$SSM
(for partial results see also
\citere{Fidalgo:2011ky}).
If these conditions are respected
at least approximately, it is possible to obtain
a particle state
at $125\gev$ with properties resembling
the ones of the SM Higgs boson without
relying on a decoupling of the
remaining doublet-like scalars
from the EW scale. In the following discussion
we will make use of the relations
\begin{equation}
Y^\nu_{ij} \ll \lambda \ , \quad
v_{iL} \ll v,v_{iR} \ , \quad
\kappa_{ijk} = \delta_{ij} \delta_{jk} \kappa_i \ .
\end{equation}
The first two inequalities
arise from the requirement to suppress
lepton-number breaking.
The third expression, i.e., 
treating $\kappa_{ijk}$
as diagonal, is
a reasonable simplification since
the superpotential can always
be written in a basis in which the right-handed
sneutrino self-couplings $\kappa_{ijk}$ are
diagonal.
The diagonal structure of $\kappa_{ijk}$ is
also preserved by the RGE
evolution of the couplings~\cite{Biekotter:2019gtq}.

The general strategy for determining the alignment
conditions follows the approach of
\citere{Carena:2015moc}, therein
applied to the NMSSM.
First, one rotates the scalar squared mass
matrix from the interaction basis into
the so-called Higgs basis, in which only
one scalar field obtains an EWSB vev.
This transformation can be expressed in
terms of a unitary transformation matrix
$U^{\rm HB}$ that will be specified below.
One can then identify relations
between the free parameters of
the model for which
the non-diagonal mass matrix
elements between the field that obtains
the EWSB vev and
the other scalar fields vanish.
The field that is aligned
with the EWSB vev then couples to the 
gauge bosons and fermions of the
SM in the same way as the
Higgs boson that is predicted
by the SM~\cite{Carena:2013ooa}.
In this limit, the CP-even Higgs sector
of the $\mu\nu$SSM consists of a
SM-like Higgs boson, $H_{\rm SM}$,
a second doublet-like \textit{heavy}
Higgs boson, $H_{\rm NSM}$, three
singlet-like Higgs bosons, $H_{1,2,3 R}$,
and three decoupled left-chiral sneutrino
states, $H_{1,2,3 L}$.\footnote{The singlet
states could still be mixed with
the second doublet-like Higgs boson $H_{\rm NSM}$.
However, assuming that the alignment
conditions are respected,
such a mixing
requires large values
of $\kappa_{i} \approx 1$
that give rise to
the presence of Landau poles below the
GUT scale. We therefore do not consider
this case here.}
We will make use of a similar unitary
transformation in terms of the matrix
$U^{\rm AB}$ for the 
CP-odd part of the Higgs sector.
This will allow us to define a tree-level mass
parameter $M_A$ in analogy to the (N)MSSM.
This mass parameter 
is roughly equal to the tree-level mass
of the massive doublet-like CP-odd Higgs boson $A$
of the $\mu\nu$SSM
in the alignment limit.

The neutral scalar mass terms in the
interaction basis are given by
\begin{equation}
\mathcal{L} = - \frac{1}{2}
\phi^T \mathcal{M}_s \phi
 - \frac{1}{2}
 \sigma^T  \mathcal{M}_p \sigma + \dots \ ,
\label{eq:masslag}
\end{equation}
where the squared mass matrices
$\mathcal{M}_s$ and $\mathcal{M}_p$
are determined by the curvature
of the scalar potential $V$ in the
direction of the fields
$\phi$ and $\sigma$
in the EW minimum.
The tree-level entries of these matrices
can be found in \citere{Biekotter:2019gtq}.
In the following it is implied
that in the mass matrices the
soft mass parameters $m_{H_d}^2$, $m_{H_u}^2$,
$(m_{\widetilde{\nu}_R}^2)_{ii}$ and
$(m_{\widetilde{\nu}_L}^2)_{ii}$
were replaced by
the vevs $v_d$, $v_u$, $v_{iR}$,
and $v_{iL}$, making use of
\refeqs{eq:tp1}--(\ref{eq:tp4}).
We transform $\mathcal{L}$ into the Higgs basis via
${\cal H}_i = U^{\rm HB}_{ij} \phi_j$
and
${\cal A}_i = U^{\rm AB}_{ij} \sigma_j$, where
\begin{equation}
\footnotesize
U^{\rm HB,AB} = 
\begin{pmatrix}
\pm v_d  / v & v_u / v & 0 & 0 & 0 &
    \pm v_{1L} / v & \pm v_{2L} / v & \pm v_{3L} / v \\
\mp v_u / v & v_d / v & 0 & 0 & 0 &
    0 & 0 & 0 \\
0 & 0 & 1 & 0 & 0 & 0 & 0 & 0 \\
0 & 0 & 0 & 1 & 0 & 0 & 0 & 0 \\
0 & 0 & 0 & 0 & 1 & 0 & 0 & 0 \\
\mp v_{1L} / v_d & 0 & 0 & 0 & 0 &
    \pm 1 & 0 & 0 \\
\mp v_{2L} / v_d & 0 & 0 & 0 & 0 &
    0 & \pm 1 & 0 \\
\mp v_{3L} / v_d & 0 & 0 & 0 & 0 &
    0 & 0 & \pm 1 \\
\end{pmatrix} \ .
\label{eq:uhbab}
\end{equation}
Here the scalar fields in the Higgs basis 
are expressed in terms of the field vector
${\cal H}^T = (H_{\rm SM}, H_{\rm NSM}, H_{1,R},
H_{2,R}, H_{3,R}, H_{1,L}, H_{2,L}, H_{3,L})$,
while the pseudoscalar
fields are expressed in terms of
${\cal A}^T = (G_0, A, A_{1,R}, A_{2,R}, A_{3,R},
A_{1,L}, A_{2,L}, A_{3,L})$, where $G_0$
is the neutral Goldstone boson and $A$ is the
doublet-like particle state.
For the elements that are given with two 
different signs, the upper and lower signs refer to
$U^{\rm HB}$ and $U^{\rm AB}$, respectively.
The different signs
arise from the fact that
the chiral superfields $H_d$
and $\widetilde{\nu}_{iL}$ have the
opposite hypercharge as $H_u$.
Thus, a field redefinition via a
global SU(2) transformation 
and a complex conjugation
of the doublet fields $H_d$
and $\widetilde{\nu}_{iL}$
is applied in order to use a basis in
which all fields with non-zero
hypercharge have the same value
of the hypercharge.
For the imaginary components of the
scalar fields, $\sigma_i$, (see the definition
as specified
in \refeq{vevcompo} and the text below) the
complex conjugation
translates into the differences of the
signs of the elements of $U^{\rm AB}$
and $U^{\rm HB}$.\footnote{One
could also redefine $H_u$ and
change the sign of only the elements
of the second column of $U^{\rm AB}$. However,
we prefer to follow the definitions of
\citere{Carena:2015moc}
to simplify the comparison with the NMSSM.}
Inverting the transformations defined
in \refeq{eq:uhbab},
one can replace $\phi_i
= (U^{\rm HB})^{-1}_{ij} {\cal H}_j$ and $\sigma_i
= (U^{\rm AB})^{-1}_{ij} {\cal A}_j$
in \refeq{eq:masslag}
to obtain the mass matrices in the
Higgs basis, such that
\begin{equation}
\mathcal{L} = - \frac{1}{2}
{\cal H}^T \mathcal{M}^{\rm HB} {\cal H}
 - \frac{1}{2} {\cal A}^T 
\mathcal{M}^{\rm AB} {\cal A} + \dots \ .
\end{equation}
As already mentioned before,
the alignment conditions are precisely
defined by demanding that the
state $H_{\rm SM}$ is aligned with the vacuum expectation value and has vanishing mixing 
with the other states, i.e., 
$\mathcal{M}^{\rm HB}_{1i} = 0$, where 
$i \neq 1$. The
resulting conditions on the
model parameters will be discussed
in the following.
In order to make a distinction between
the fields in the Higgs basis $H_i$
and the (loop-corrected) mass eigenstates,
we will use the notation $h_i$ for the latter,
where the index $i$ reflects the mass
hierarchy of the scalars.

\subsection{First alignment condition}
The first alignment condition
results from requiring that the
mixing between the two Higgs
doublet states $H_d$ and $H_u$ is such that
(ignoring for now a possible singlet
admixture)
one mass eigenstate $h_i \approx H_{\rm SM}$
couples to the SM gauge bosons with a coupling
that is approximately equal to the one of the 
SM Higgs boson, and the other
doublet-like state $h_j \approx H_{\rm NSM}$
has a vanishing coupling to gauge bosons
as it does not acquire an EWSB vev.
The alignment condition arises from the
requirement
\begin{equation}
\mathcal{M}^{\rm HB}_{12} = \mathcal{M}^{\rm HB}_{21} =
- \overline{M}_Z^2 c_{2 \beta} s_{2 \beta}  = 0 \ ,
\label{al1origin}
\end{equation}
where $\overline{M}_Z^2 =
M_Z^2 - v^2 \lambda_i \lambda_i / 2$
with the $Z$ boson mass, $M_Z \approx 91\gev$,
summation over $i=1,2,3$ is implied,
and we use the short-hand notation
$c_x \equiv \cos x$ and $s_x \equiv \sin x$.
Using also that
\begin{equation}
\mathcal{M}^{\rm HB}_{11} =
\overline{M}_Z^2 c_{2 \beta}^2 +
\frac{1}{2}
\lambda_i \lambda_i
v^2 \ ,
\label{eqmhsmtl}
\end{equation}
it is convenient to write
(with $t_x = \tan x$)
\begin{align}
\mathcal{M}^{\rm HB}_{12} - \mathcal{M}^{\rm HB}_{11}
 / t_\beta =
- c_{2 \beta} M_Z^2 / t_\beta
- c_\beta s_\beta v^2 \lambda_i \lambda_i
\ .
\label{al1relat}
\end{align}
The expression shown in \refeq{eqmhsmtl}
is the squared tree-level mass of
the SM-like Higgs boson in the exact
alignment limit. It is known that
in SUSY models
large radiative corrections from the
stop sector can be
present~\cite{Slavich:2020zjv}.
As discussed in
\citere{Carena:2015moc},
to a very good approximation
these radiative corrections
only enter in $\mathcal{M}^{\rm HB}_{11}$,
but not in the off-diagonal
entry $\mathcal{M}^{\rm HB}_{12}$.
Thus, in order to account for those radiative
corrections it is sufficient to substitute
${\mathcal{M}^{\rm HB}_{11} \rightarrow
m_{h_{\rm SM}}^2 \approx (125\gev)^2}$
in \refeq{al1relat}, and
the condition shown in \refeq{al1origin}
can be written as
\begin{equation}
\lambda_1^2 +
\lambda_2^2 +
\lambda_3^2 =
\frac{m_{h_{\rm SM}}^2 -
    M_Z^2 c_{2 \beta}}
{v^2 s_\beta^2} \ .
\label{eq:alcond1}
\end{equation}
This is what we will refer to in the following
as the
first alignment condition, which is
a straightforward generalization of
the NMSSM condition~\cite{Carena:2015moc}.

\subsection{Second, third and fourth
alignment conditions}
\label{sec:234conds}
In the NMSSM, there is a second alignment
condition that ensures that there is no
mixing between $H_{\rm SM}$ and the
gauge singlet field. In the $\mu\nu$SSM,
there are three gauge singlet right-handed
sneutrino fields. Each of the
them can potentially
mix with the SM-like Higgs boson.\footnote{On
the other hand, the mixing of the doublet fields
$H_{d,u}$ with the left-handed sneutrino fields
$\widetilde{\nu}_{iL}$
is always suppressed by the tiny values of
$Y^\nu_{ij}$ and $v_{iL}$,
such that automatically
$\mathcal{M}_{16}^{\rm HB} \approx 0$,
$\mathcal{M}_{17}^{\rm HB} \approx 0$ and
$\mathcal{M}_{18}^{\rm HB} \approx 0$.}
In order to
avoid a singlet admixture in the state
$H_{\rm SM}$,
one has to fulfill three more conditions, i.e.,
\begin{equation}
\mathcal{M}^{\rm HB}_{13} = 0 \ , \quad
\mathcal{M}^{\rm HB}_{14} = 0 \ , \quad
\mathcal{M}^{\rm HB}_{15} = 0 \ .
\end{equation}
Thus, alignment without decoupling is obtained if these 
three conditions are fulfilled together with the condition
of \refeq{eq:alcond1}.
Inserting the mass matrix elements, one finds
the conditions
\begin{align}
A^\lambda_i =
\frac{\mu}{c_\beta s_\beta} -
\sqrt{2} \kappa_{i} v_{iR}
\ , \quad  i = 1,2,3 \ .
\label{alcond234}
\end{align}
The corresponding NMSSM condition was
derived in \citere{Ellwanger:2006rm},
and it was already
generalized to the $\mu\nu$SSM
in \citere{Fidalgo:2011ky}. However, a phenomenological
exploration of the $\mu\nu$SSM parameter
space in combination with the first
alignment condition shown in \refeq{eq:alcond1}
has not been carried out yet.
One can also obtain a relation in closer
analogy to the NMSSM by realizing that,
if all three conditions shown above
are fulfilled, then also the sum
vanishes, such that
\begin{equation}
\frac{1}{v} \left(
v_{1R} \mathcal{M}^{\rm HB}_{13} +
v_{2R} \mathcal{M}^{\rm HB}_{14} +
v_{3R} \mathcal{M}^{\rm HB}_{15}
\right) = 0 \ .
\end{equation}
This expression allows 
one to replace
the terms $\sim A^\lambda_i =
T^\lambda_i / \lambda_i$ 
by the tree-level mass parameter
of the 
doublet-like CP-odd Higgs boson
given by
\begin{align}
M_A^2 = \mathcal{M}^{\rm AB}_{22} =
\frac{1}{2 c_\beta s_\beta}
\left(
\sqrt{2}
T^\lambda_i v_{iR} +
\kappa_{i} \lambda_i v_{iR}^2
\right) \ ,
\end{align}
where the summation over $i=1,2,3$ is
implied and, as mentioned in the beginning,
terms $\sim v_{iL}$ are not written out.
This leads to the condition
\begin{equation}
\frac{M_A^2 s_{2 \beta}^2}{4 \mu^2} +
\frac{s_{2 \beta}}{2}
\frac{\kappa_{i} \lambda_i v_{iR}^2}
{2 \mu^2}
= 1 \ ,
\label{notal2}
\end{equation}
where again the summation over $i=1,2,3$ is
implied.
The first term in \refeq{notal2} exactly coincides with
the NMSSM result~\cite{Carena:2015moc}.
The second term has
become considerably more complicated
due to the presence of more than
one gauge singlet field.
However, assuming that two of the three
singlets are decoupled, e.g.,
$\lambda_{2,3} \rightarrow 0$,
the NMSSM formula is recovered when
substituting $v_{1R}^2 = 2 \mu^2 /
\lambda_1^2$ in the surviving term.
It should be noted, however, that
in the $\mu\nu$SSM the condition
shown in \refeq{notal2} is a necessary,
but not a sufficient
criterion for achieving the alignment limit.
In other words, \refeq{notal2} must be true in the
alignment limit,
i.e., when the conditions
shown in \refeq{alcond234} are fulfilled
separately,
but not all parameter
configurations that respect \refeq{notal2}
correspond to the alignment limit.
Finally,
by noting that in order to
fulfill the first alignment
condition one has to require rather large
values of ${\lambda_1^2 + \lambda_2^2 +
\lambda_3^2 \gtrsim 0.6^2}$,
a possibility to avoid Landau
poles below the GUT scale is to
assume $\kappa_{i}
\ll \lambda_i$.  In this case the
second term in \refeq{notal2} is small,
and it is
sufficient to 
use as a condition
$M_A^2 s_{2 \beta}^2 / (4 \mu^2) \approx 1$,
just as in the NMSSM~\cite{Carena:2015moc}.

\section{Vacuum stability}
\label{vacstab}
The main goal of this work is
to investigate the stability
of the EW vacuum of the
$\mu\nu$SSM at temperature $T=0$.
This analysis can be divided into
two different tasks. First, one has
to check whether there are 
local minima present
in the scalar potential that are
deeper than the EW minimum. If there is
no such unphysical minimum, the EW
minimum is the global one, and the
EW vacuum is
stable (see, however, the discussion 
on vacuum trapping
in \refse{intromunu}). If one or more deeper
unphysical minima are present, one has to 
investigate in
a second step whether the EW vacuum
can be regarded as sufficiently
long-lived or whether it would rapidly
decay into one of the unphysical vacua.
In \refse{minimization} we will briefly
introduce our approach for determining
the different minima of the neutral
scalar potential. Subsequently, we
give details on the calculation of the
lifetime of metastable EW vacua
in \refse{bouncer}.

The computations that are carried out for testing
vacuum stability
as described in the following are
implemented in the public code
\texttt{munuSSM v.1.1.0}~\cite{Biekotter:2017xmf,
Biekotter:2019gtq,Biekotter:2020ehh}. More details
on the implementation and simple user
instructions are given in
\refap{appcode}.

\subsection{Finding potentially
dangerous unphysical vacua}
\label{minimization}
Due to the large hierarchy between
the doublet and singlet vevs,
$v_d,v_u,v_{iR} \approx v$, and
the left-handed sneutrino vevs,
$v_{iL} \approx Y^\nu_{ij} v$, it is
very challenging to find the local
minima of the potential $V$ with the help
of usual minimization algorithms.
In particular, gradient based methods
poorly converge due to the nearly flat directions
in the potential 
originating from the large
hierarchy of parameters. Moreover,
such algorithms
crucially depend on the initial
conditions (such as the starting values
for the fields), and
it is very challenging to find all (or at least
most) of the various different minima of
the $\mu\nu$SSM scalar potential
in this way.

We therefore 
use an approach that is based on directly solving
the polynomial minimization conditions
shown in \refeqs{eq:tp1}\nobreakdash--(\ref{eq:tp4}).
The solutions yield
the stationary points of
the potential. 
In a second step 
we determine the stationary points corresponding
to a local
minimum of the potential by calculating the
eigenvalues of the Hessian matrix of the
potential for each solution that was found.
For all local minima 
the value of the
scalar potential is calculated. In this way the minima 
that are deeper than the EW minimum,
and therefore potentially dangerous for
the stability of the EW vacuum, are determined. 
For each minimum of this kind
a calculation of the transition probability
is carried out as described in
\refse{bouncer}.

The solutions to the stationary conditions
were obtained using the public code
\texttt{HOM4PS-2.0}~\cite{Lee:hom4ps2}, which is an
implementation of
the polyhedral homotopy continuation
method~\cite{Bernstein:1975,10.2307/2153370,Li:2003hp}.
In general, the code is efficient 
in finding all existing solutions. 
In our analysis it happened only in rare cases 
that a solution was missed by \texttt{HOM4PS-2.0} (see also the 
discussion in \refse{numana} below), which
could be tested by computing the same parameter
point several times and comparing the number of solutions
that were found.
This can
have different reasons. First, it can happen that
the code converges
twice to the same solution 
from two different
starting points on the unit circle
in the complex plane,
for instance, if
at intermediate stages of the
algorithm two different solutions
only differ by the tiny
values of the left-handed
sneutrino fields
$\widetilde{\nu}_{iL}^{\mathcal{R}}$.
Secondly, one has to define
a numerical uncertainty for the imaginary parts
of the  solutions to the stationary conditions
in order to identify which of the solutions
correspond to a real solution, since they are the
ones we are interested in here.
If this uncertainty is set to too small
values, i.e., below the numerical precision
of the algorithm, a real solution
is misidentified as a complex one and
dismissed erroneously (see also
\citere{Hollik:2018wrr}).

We finally remark that it is important to
remove the redundant degrees of freedom
related to the gauge symmetries. Otherwise,
any stationary point would be transformed
into flat directions along the redundant
degrees of freedom, and the method introduced
above would fail. In our analysis, this
problem is absent since we restrict the
analysis to the case in which only the
CP conserving real parts of the neutral
scalar fields are allowed to develop a
vev, such that no degeneracies related
to the gauge symmetries are present.

\subsection{
Lifetime of 
the EW vacuum 
in the presence of deeper minima}
\label{bouncer}
In case one or more deeper
unphysical minima are
present, the EW vacuum 
could decay via quantum tunneling effects
into an unphysical vacuum. The probability for
this vacuum decay mainly depends
on the euclidean bounce action $S_E$ for
the classical path in field space
from the EW into the unphysical minimum.
The computation of this
bounce action is a complicated
task, in particular for cases where 
the number of
field dimensions is quite large.\footnote{See,
for instance, \citere{Athron:2019nbd}
for more details and different numerical
approaches applied in the literature
to compute the bounce action.}
Due to the fact that in the $\mu\nu$SSM
there are (at least) eight real fields
that have to be considered, and since
there are large hierarchies between the field
values, one has to find a numerically
efficient, fast and 
reliable way of calculating
the tunneling probabilities. We 
follow here
the approach of \citere{Hollik:2018wrr}
(see also \citeres{Hollik:2016dcm,
Hollik:2018nhq,Ferreira:2019iqb}), in which $S_E$
is obtained by making use of a semi-analytical
approximation, and 
a simple criterion
is applied in order to 
determine whether a certain
parameter point 
features
a short-lived,
long-lived, or stable EW vacuum.
We briefly describe this
approach in the following.

At tree level the CP-even scalar Potential $V(\phi_i)$
as given in \refeq{eq:fullpot}
can be brought into the form~\cite{Hollik:2018wrr}
\begin{equation}
V_{\rm EW}(\varphi, \hat{\varphi}) =
\lambda(\hat\varphi) \varphi^4 -
A(\hat\varphi) \varphi^3 +
m^2(\hat\varphi) \varphi^2 \ ,
\label{eq1dpot}
\end{equation}
where the fields were redefined via a shift
to the EW vacuum configuration, such that
$\phi_i = \phi^{\rm EW}_i + \varphi_i$,
and the field vector $\varphi_i$
is then expressed in terms
of its norm $\varphi = (\varphi_i
\varphi_i)^{1/2}$ and the unit vector
$\hat\varphi_i$, such that
$\varphi_i = \varphi \hat\varphi_i$.
Since by construction
$\phi^{\rm EW}$ solves the stationary conditions
of the potential there is no linear term
in \refeq{eq1dpot}.
In addition, the field independent terms
were discarded in \refeq{eq1dpot}, so that
$V_{\rm EW}(0,\hat\varphi) = 0$.

After having identified the unphysical
minima, one can investigate $V_{\rm EW}$
along the direction from the EW minimum
towards each of the unphysical minima 
using a straight-path approximation~\cite{Hollik:2018wrr}.
For a fixed unit vector $\hat\varphi_i \equiv \hat
\varphi_{i,d}$ pointing towards
a deeper unphysical minimum
$\phi_{i,d}$, the coefficients
$\lambda(\hat\varphi_{i,d})$, $A(\hat\varphi_{i,d})$
and $m^2(\hat\varphi_{i,d})$ are positive, and
the function $V_{\rm EW}(\varphi) \equiv
V_{\rm EW}(\varphi, \hat{\varphi} = \hat
\varphi_{i,d})$ has
two local minima, the EW minimum at
$\varphi = 0$ and the unphysical minimum
at $\varphi_d \equiv \Delta \phi$, which
are separated by a potential barrier
originating from the trilinear $A$--term
(see also \reffi{fig:potew} in the numerical
discussion below). The unphysical minimum at $\phi_{i,d}$
is therefore the global minimum
of $V_{\rm EW}(\varphi)$, and
$V_{\rm EW}(\varphi_d)$
is equal to 
the potential
difference $\Delta V_{\rm EW}$ between the
unphysical minimum and the EW minimum.

For the one-dimensional potential
$V_{\rm EW}(\varphi)$, which has been obtained as
described above making use of a straight-path approximation, 
it is possible to calculate
the bounce action for a transition from the
EW into the deeper unphysical minimum.
This calculation was carried out in
\citere{Adams:1993zs} for potentials of the shape
as defined in \refeq{eq1dpot}, and it was shown
that a very good numerical fit to the
result for $S_E$ is given by
\begin{equation}
S_E = \frac{\pi^2}{3 \lambda}
\left( 2 - \delta \right)^{-3}
\left(13.832 \delta -
10.819 \delta^2 +
2.0765 \delta^3 \right) \ ,
\quad \delta =
\frac{8 \lambda m^2}{A^2} \ .
\label{eq:SE}
\end{equation}
In our numerical analysis we compared
the results for $S_E$ based on this
semi-classical approximation with
the ones obtained with the public code
\texttt{cosmoTransitions}~\cite{Wainwright:2011kj},
and found deviations between both
calculations below $1\%$ for all parameter
points considered.\footnote{For this
comparison, we implemented the one-dimensional
scalar potential $V_{\rm EW}(\varphi)$ into
\texttt{cosmoTransitions}. Thus, the calculation
does not benefit from the path deformation
method as provided by the code for the case
of multiple field dimensions. Given the full
CP-even scalar potential $V$ of the $\mu\nu$SSM, 
this method could in principle be applied
in order to obtain a more precise result.
However, we observed that the path deformation
algorithm as implemented in
\texttt{cosmoTransitions} did not
converge successfully.
We attribute this to the large number of fields
and the large hierarchy of the field values
in the~$\mu\nu$SSM.}
If the kinetic terms are canonically
normalized, the decay rate $\Gamma$ for the quantum
tunneling into the unphysical minimum
under consideration per space volume $V$
is then given by~\cite{Coleman:1977py,Callan:1977pt}
\begin{equation}
\frac{\Gamma}{V} =
K \mathrm{e}^{- S_E} \ ,
\label{eq:decayrate}
\end{equation}
where $K$ is a sub-leading prefactor
that incorporates higher-order corrections.
For practical purposes, it appears to
be sufficient to estimate 
$K$ only based on dimensional
arguments (see the discussion in \citere{Hollik:2018wrr}), 
such that $K \approx M^4$. Here
$M$ represents a typical mass scale
of the model. In order to estimate whether
the EW vacuum of a
parameter space point of the $\mu\nu$SSM
is sufficiently long-lived, or
whether it would rapidly decay
into one of the unphysical vacua, we
followed the criteria formulated in
\citere{Hollik:2018wrr}. A parameter point
is considered to feature a
long-lived 
EW vacuum
if the values for $S_E$ for all possible
transitions into
unphysical vacua are larger than $440$.
On the contrary, if for one of the possible
transitions we find $S_E < 390$, the
EW vacuum is considered to
be short-lived compared to the
lifetime of the universe, and the
corresponding parameter point
should be rejected.
The region in between, $390 < S_E < 440$, 
arises from the uncertainty stemming from the
unknown prefactor $K$.
In our numerical analysis we will
separately display parameter 
points for which $S_E$ falls into this region 
where in our approach no clear distinction 
between a short-lived and a
sufficiently long-lived EW vacuum is possible.

Before starting the numerical discussion
in \refse{numana},
we point out
that in the $\mu\nu$SSM
more than one transition into different
unphysical minima with $S_E < 440$ can be
present for a single parameter point.
Here it should be noted that the lifetime of
the EW vacuum is practically unaffected by the
number of deeper unphysical minima
into which the EW vacuum could decay, because
the numerically dominant
exponential term $\mathrm{e}^{-S_E}$ is determined
based on each possible vacuum decay individually.
Taking into account that the prefactor
$K$ in our approach
is estimated only based on dimensional
arguments,
it is therefore sufficient to equate the
total decay rate of the EW vacuum with
the maximum of the individual decay rates
belonging to each possible transition.
Consequently, our above considerations
remain valid also in the presence
of several unphysical vacua.

\section{Numerical analysis}
\label{numana}
As already mentioned in \refse{intro},
in our analysis of the possible impact of EW
vacuum stability constraints
on the parameter space of the $\mu\nu$SSM we will focus on the 
alignment without decoupling region.
In the $\mu\nu$SSM an important difference to the
corresponding limit in the NMSSM arises
from the fact that there are three gauge singlet
scalar fields present in the $\mu\nu$SSM.
Each of these singlet sneutrino fields
is coupled to the Higgs doublet fields via
a portal coupling $\lambda_i$, whereas
in the NMSSM there is only one such
coupling $\lambda$. Consequently,
while in the NMSSM the
first alignment condition fixes the
value $\lambda$ as a function of
$\tan\beta$, in the $\mu\nu$SSM
the corresponding condition as given in
\refeq{eq:alcond1} can be fulfilled for different
individual values of $\lambda_i$, while
only their squared sum is given as a
function of $\tan\beta$.
In order to be as general as possible, we decided
to choose different values for $\lambda_i$ in a
hierarchical structure, such that one
right-handed sneutrino is rather strongly coupled
to the Higgs doublet fields, another one
is moderately coupled, and the third one
is largely decoupled. The precise values of
the free parameters are summarized in
\refta{tab:bp2pa}, and we will briefly discuss
the choice of their values
in the following.

\begin{table}
\centering
{\renewcommand{\arraystretch}{1.4}
\footnotesize
\begin{tabular}{cccccccccc}
$\tan\beta$ & $v_{iL}$ & $v_{iR}$ & $\lambda_1$ &
  $\lambda_2$ & $\lambda_3$ & $\kappa_1$ & $\kappa_2$ &
    $\kappa_3$ & $Y^\nu_i$ \\
\hline
$2$ & $1\cdot 10^{-4}$ & $[770,840]$ & $0.54$ & $0.36$ &
  $0.005$ & $0.41$ & $0.42$ & $0.43$ & $1.7\cdot 10^{-7}$ \\
\hline
\hline
$M_S$ & $A^t$ & $A^\kappa_i$ & $A^\nu_i$ &
  $A^\lambda_i$ & $A^\lambda_2$ &
    $A^\lambda_3$ & $M_1$ & $M_2$ & $M_3$ \\
\hline
$2500$ & $500$ & $[-915,-800]$ & $-1500$ & $816$ & $805$ & $794$ &
  $200$ & $400$ & $2000$
\end{tabular}
}
\caption{\small Parameter values for the scenarios
discussed in the numerical analysis.
Dimensionful parameters are given
in GeV. For the scan over $A^\kappa_i$ we
fixed $v_{iR} = 800\gev$. For
the scan over $v_{iR}$ we fixed $A^\kappa_i =
-880\gev$.}
\label{tab:bp2pa}
\end{table}

We have chosen
$\tan\beta = 2$, such
that according to \refeq{eq:alcond1}
the $\lambda_i$ have to fulfill
the constraint $\sqrt{\lambda_1^2 +
\lambda_2^2 + \lambda_3^2}
\approx 0.65$. Taking into account the
above, we use values of $\lambda_1 = 0.54$,
$\lambda_2 = 0.36$ and $\lambda_3 = 0.005$.
These values yield a relatively large
enhancement of the tree level mass of the
SM-like Higgs boson. Thus, no large radiative
corrections are required to achieve a
physical mass of $\approx 125\gev$, and the
values for the stop mass parameters,
$m_{\widetilde{Q},\widetilde{t}}
= M_S = 2.5\tev$, the soft trilinear
stop coupling, $A_t = 500\gev$, and
the gluino mass parameter, $M_3=2\tev$, are
set accordingly.\footnote{All soft slepton
and squark mass parameters were set to
be equal to $M_S$,
and the remaining soft trilinear couplings
of the sleptons and squarks are set to zero.
For the (tree-level) analysis of the vacuum stability
these values have no relevance.}
In order to fulfill the remaining alignment
conditions as given in \refeq{alcond234}, we
set the values of $A^\lambda_i$
and $\kappa_i$ as given
in \refta{tab:bp2pa}. The parameters $v_{iR}$ and
$A^\kappa_i$ were varied in a range that
yield right-handed sneutrino masses in
the vicinity of $125\gev$, since this
is the mass scale at which the sneutrinos
can be mixed sizably with the
SM-like Higgs boson.
The corresponding parameter space
is also of phenomenological 
interest as it
can be probed experimentally
by the LHC and possible future
colliders~\cite{Biekotter:2019gtq}.
As we will discuss below, the
parameters $v_{iR}$ and $A^\kappa_i$
are particularly important
for the presence of a potential barrier
between the EW minimum and deeper
unphysical minima. Therefore, also the
decay rates of the EW vacuum strongly
depend on these parameters.
Combining the impact of the sneutrino
mixing with the SM-like Higgs boson
on the one hand, and the considerations
regarding the EW vacuum stability as
discussed below on the other hand,
there is an interesting complementary
between experimental
and theoretical constraints 
on the parameter
space under investigation.

While we choose the same values of the parameters
$v_{iR}$, $A^\nu_i$ and $A^\kappa_i$
for the three generations, i.e.\ 
$v_R \equiv v_{iR}$, 
$A^\nu \equiv A^\nu_i$ and
$A^\kappa \equiv A^\kappa_i$,
we set the values of
$\kappa_i$ slightly different.
These differences are commonly
included in studies of the $\mu\nu$SSM,
since they ensure that there
are not two right-handed
sneutrinos with almost exactly the same
masses in case the values for $\lambda_i$
are chosen to be equal,
due to an accidental symmetry in the
scalar potential (see
\citeres{Ghosh:2014ida,Biekotter:2019gtq} for details).
Even though the $\lambda_i$ are not
uniform here,
we still follow this approach in order
to increase the mass differences
between the three families of
right-handed sneutrinos slightly more.
This leads to the fact that also the values
of $A^\lambda_i$ are not uniform,
as their values are fixed by the second,
third and fourth alignment conditions
shown in \refeq{alcond234}.

The remaining parameters
are mostly related to the
EW seesaw mechanism
and not
directly relevant for the discussion here.
In particular, for $v_{iL} = 10^{-4}\gev$
and $Y^\nu_{ij} = \delta_{ij} Y^\nu_i =
1.7\cdot 10^{-7}$ one finds left-handed
neutrino masses 
at the sub-eV
level.
Note, however, that we
did not include a precise fit to the
neutrino oscillation data in terms of
squared mass differences and mixing angles,
as such details have no impact
on our discussion. Finally, the gaugino
mass parameters are set to $M_1 = 200\gev$
and $M_2 = 400\gev$, such that they are of
roughly the same order as $\mu \approx 500\gev$
and the left-handed sneutrino masses
$m_{\widetilde{\nu}_{iL}} \approx 560\gev$,
where $m_{\widetilde{\nu}_{iL}}$
are controlled by the
values of $A^\nu_i = -1.5\tev.$
Since the left-handed sneutrinos are
substantially heavier than the right-handed
sneutrinos and the SM-like Higgs boson,
and since they are practically
not mixed with each other, we do not discuss
the left-handed sneutrinos any further
in the following.

As shown in \refta{tab:bp2pa}, we vary the two
parameters $v_R \equiv v_{iR}$ and
$A^\kappa \equiv A^\kappa_i$.
We start our numerical investigation by varying each of the
two parameters individually and analyze
how the EW vacuum stability depends on
them. At the end of this section,
we will summarize our numerical discussion
by showing the results in the two-dimensional
parameter plane ${v_R\text{--}A^\kappa}$ arising
from a grid scan over both parameters.
In the first scan, we
set $v_R = 800\gev$ and vary $A^\kappa$.
In the second scan, we set $A^\kappa = -880\gev$
while varying $v_R$ in the given range.
For each parameter point, we first calculated
the radiatively corrected Higgs boson spectrum.
Afterwards, we checked the point against constraints
from BSM Higgs boson searches and the signal
rate measurements of the Higgs boson at
$\approx 125\gev$.
For the parameter points that passed the
constraints, we determined
the unphysical minima
and finally calculated the bounce actions for
the available transitions from the EW
vacuum into all potentially
dangerous unphysical vacua.
The whole analysis chain can be performed
with the public code
\texttt{munuSSM v.1.1.0}~\cite{Biekotter:2017xmf,
Biekotter:2019gtq,Biekotter:2020ehh},
utilizing the interfaces to the public codes
\texttt{FeynHiggs v.2.16.1}~\cite{Heinemeyer:1998yj,
Heinemeyer:1998np,Degrassi:2002fi,Frank:2006yh,
Hahn:2013ria,Bahl:2016brp,Bahl:2017aev,
Bahl:2018qog},
\texttt{HiggsBounds v.5}~\cite{Bechtle:2008jh,
Bechtle:2011sb,Bechtle:2013gu,Bechtle:2013wla,
Bechtle:2015pma,Bechtle:2020pkv},
\texttt{HiggsSignals v.2}~\cite{Bechtle:2013xfa,
Stal:2013hwa,Bechtle:2014ewa,Bechtle:2020uwn}
and \texttt{HOM4PS-2.0}~\cite{Lee:hom4ps2}.
For the investigation of the stability of
the EW vacuum, we have extended the code
\texttt{munuSSM}
by incorporating the new
subpackage \texttt{vacuumStability}.\footnote{The
user instructions to reproduce our
numerical analysis, in particular
the ones related to the new vacuum
stability feature,
are summarized in \refap{appcode}.}

\begin{figure}
\centering
\includegraphics[width=0.44\textwidth]{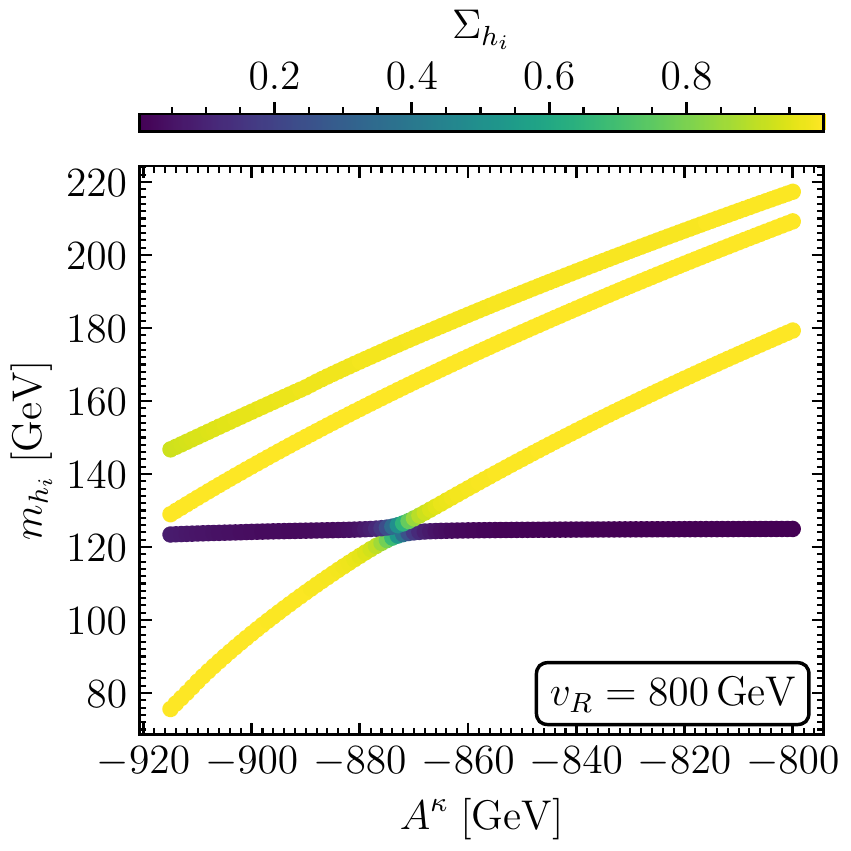}~
\includegraphics[width=0.44\textwidth]{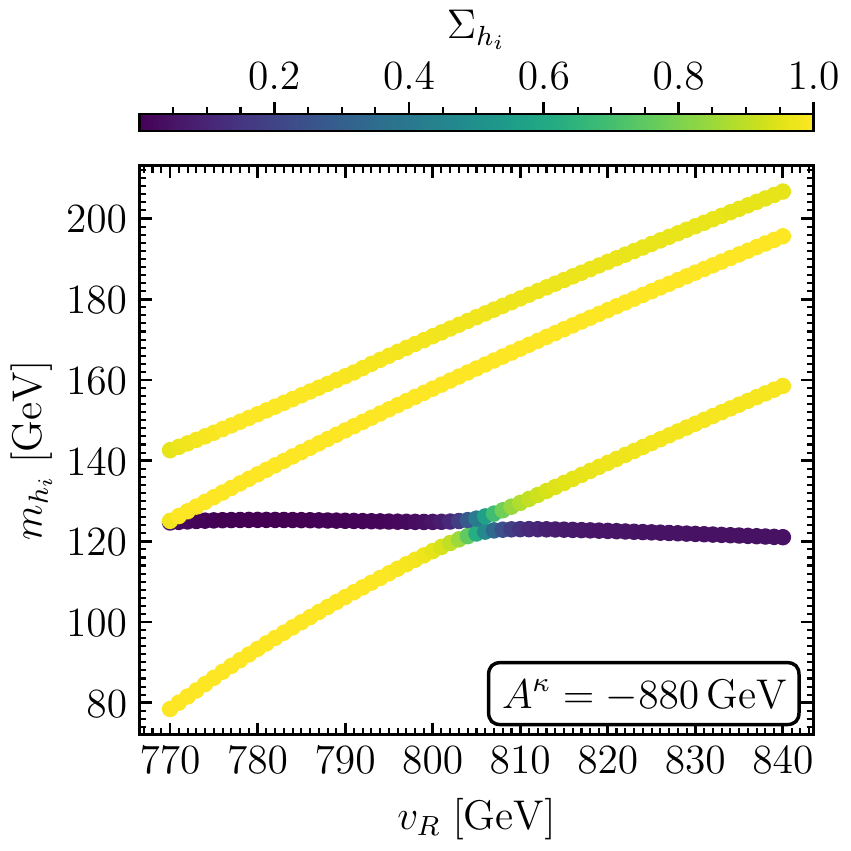}
\caption{\small 
Masses of the Higgs bosons that are in the vicinity
of $125\gev$
as function of $A^\kappa$ (left) and $v_R$ (right). 
All points pass the \texttt{HiggsBounds}
test. All points except the ones with $m_{h_1} \approx
m_{h_2} \approx 125\gev$ pass the \texttt{HiggsSignals}
test (see text). The color coding indicates the gauge 
singlet component 
for each of the displayed Higgs bosons.}
\label{fig:spec}
\end{figure}

In \reffi{fig:spec} we show the radiatively
corrected Higgs boson masses $m_{h_{1,...4}}$,
corresponding to a SM-like Higgs boson
at $\approx 125\gev$ ($\equiv h_{125}$)
and the three CP-even right-handed
sneutrino states with masses not far
below or above $125\gev$. The colors of the
points indicate the gauge singlet component
of each Higgs boson, which we define as
\begin{equation}
\Sigma_{h_i} = 
\sum_{j=3}^5 \left|
R^H_{ij} \right|^2 \ ,
\end{equation}
where $R^H$ is the loop corrected
mixing matrix that transforms the
CP-even fields $\phi_i$ from the interaction
basis 
into the fields $h_i$ in
the mass eigenstate basis.\footnote{The
radiative corrections to the mass
matrix of the neutral scalars are evaluated
including the momentum dependence, such
that $R^H$ is not a
unitary matrix~\cite{Biekotter:2019gtq}.
However, the non-unitary pieces are
numerically tiny and not relevant
for our discussion.}
One can see that, as expected from
the discussion in \refse{sec:234conds}, $h_{125}$
has only a small singlet component over
almost the entire scan range, despite the
fact that the right-handed sneutrinos are
close in mass. The only exceptions are the
points for which $h_{125}$ is practically
degenerate with the lightest right-handed
sneutrino and where a ``level-crossing'' occurs, 
i.e.\ $h_{125}$ changes its character from being the next-to-lightest 
state, $h_2 = h_{125}$, to the lightest state, $h_1 = h_{125}$. 
In the region where the level-crossing takes place the properties 
of a SM-like Higgs boson are in fact shared between the states
$h_1$ and $h_2$, and
both Higgs bosons contribute to the signal
rate measurements at the LHC for the particle state
at $\approx 125\gev$.

Using \texttt{HiggsSignals}
we applied a $\chi^2$-test
to each parameter point shown in
\reffi{fig:spec} in order to compare
the predicted signal rates of $h_{125}$
with the experimental measurements.\footnote{We
assumed a theoretical uncertainty of $3\gev$
for the calculation of the masses of the Higgs bosons.}
We discuss the results of this test in terms
of $\Delta\chi^2 = \chi^2_{\mu\nu{\rm SSM}} -
\chi^2_{\rm SM}$, where $\chi^2_{\mu\nu{\rm SSM}}$
is the fit result of the parameter points
and $\chi^2_{\rm SM} = 84.4$ is the fit result
assuming a SM Higgs boson at $125\gev$.
For the points of the scan
over $A^\kappa$ (left plot
of \reffi{fig:spec}), we find
a good compatibility with the measured properties 
of $h_{125}$, resulting in $\Delta \chi^2$ values of 
$\Delta \chi^2 \approx -1$ to $3$, except for the points
for which $m_{h_1} \approx m_{h_2} \approx 125\gev$,
where
$\Delta\chi^2 \approx 4$ to $6$.
Here it should be noted that $A^\kappa$ does
not appear in the alignment conditions. 
Accordingly,
the variation of $A^\kappa$
does not have a sizable impact
on the mixing of $h_{125}$ with the
right-handed sneutrino states, except
for the degenerate region with $m_{h_1}
\approx m_{h_2}$, in which the assumptions
on which the alignment condition rely are
not fulfilled. Consequently, 
no sizable impact on
the properties of $h_{125}$ is expected
from the variation of $A^\kappa$
in the analyzed parameter space outside of the degenerate 
region.

The situation is different when $v_R$
is varied (right plot
of \reffi{fig:spec}),
because it gives rise to a variation of the
$\mu$ parameter that appears in
\refeq{alcond234}. 
The constraint $\Delta \chi^2 < 6$
was used on the scan range,
except for the points
for which $h_1$ is degenerate with $h_2$,
where we find values of $\Delta\chi^2$ of up to
$\Delta\chi^2 \approx 12$. Since we
are mainly interested in the impact of
the constraints on
vacuum stability, and since the mass-degenerate 
points do not show any 
particular
feature regarding the constraints from
vacuum stability compared
to the other points, we do not discard points
with larger values of $\Delta \chi^2$ in the regions
with $m_{h_1} \approx m_{h_2} \approx 125\gev$
(we note that in this region the predicted signal rates,
to which both $h_1$ and $h_2$ contribute, show sizable deviations 
from the
measured values).
For the upper end of 
the displayed $v_R$ range 
$\Delta\chi^2$ increases as a consequence of the 
decrease in the predicted value for the mass
of the SM-like Higgs boson.
For values of $v_R \gtrsim 830\gev$ we
find $m_{h_{125}} \lesssim 122\gev$, 
which is outside of the allowed range that we have employed 
based on a theoretical uncertainty of
the Higgs-mass prediction
of $3\gev$. The lower end of the scan
range of $v_R$ arises from constraints from
direct searches, as described below.

Regarding the \texttt{HiggsBounds} test,
which confronts each parameter point
with the available constraints from
BSM Higgs-boson searches at colliders,
all points of the scan over $A^\kappa$ pass
the experimental constraints. This is related
to the fact that the particles corresponding
to the right-handed sneutrinos are very singlet-like 
and have strongly suppressed couplings to
the SM particles.\footnote{Currently
\texttt{HiggsBounds} does not take into account
the pair production of charged scalars. Thus, the
production of the neutral sneutrinos via
slepton decays is also not taken into account
here.
However, the decays of the sleptons into
the gauge singlet sneutrinos plus another
slepton are highly suppressed and not relevant
at the LHC.}
Only for the smallest values
of $A^\kappa$ one can see an
increase of the doublet
component of the heaviest right-handed
sneutrino ($h_4$), such that the channel
$p p \to h_4 \to Z Z^*$ becomes
important, which was searched for by CMS
in the two-lepton and four-lepton
final states and including width
effects~\cite{Sirunyan:2018qlb}.
However, the predicted cross sections remain
roughly $20\%$ below the experimental limits even
in this range of $A^\kappa$.
The same experimental channel
is also the most sensitive one in the scan
over $v_R$, yielding an exclusion for
values of $v_R < 770\gev$.
Accordingly, we use this value as the
lower limit of the scan over $v_R$.

\begin{figure}
\centering
\includegraphics[height=7.2cm]{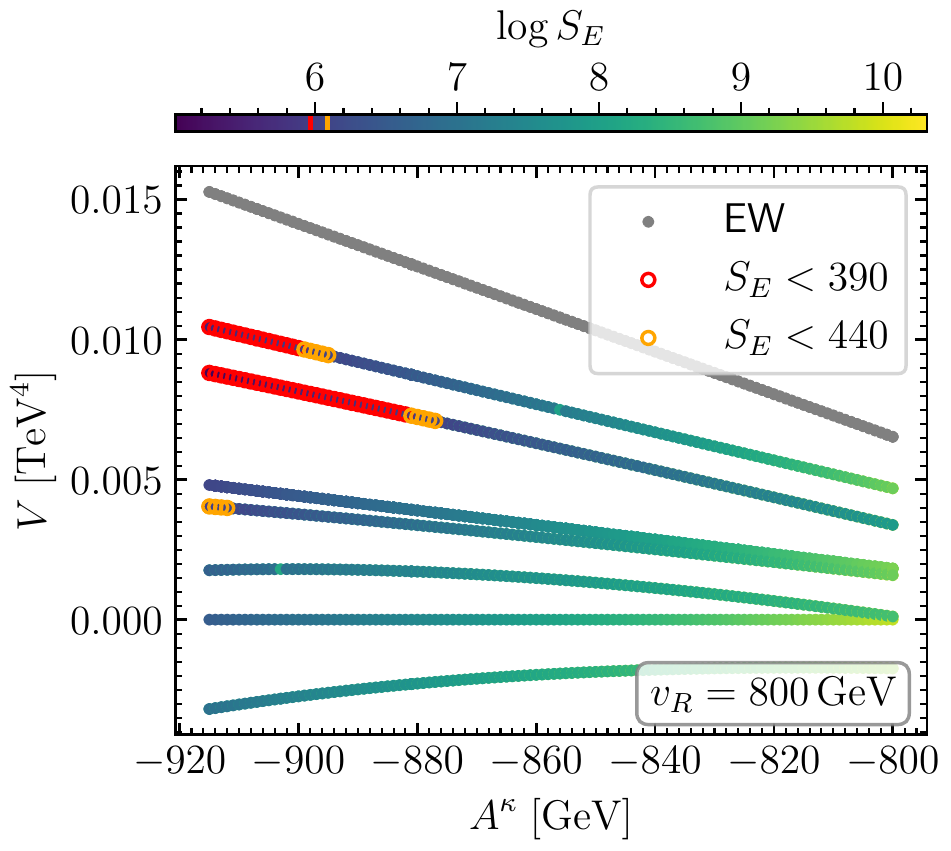}~
\includegraphics[height=7.2cm]{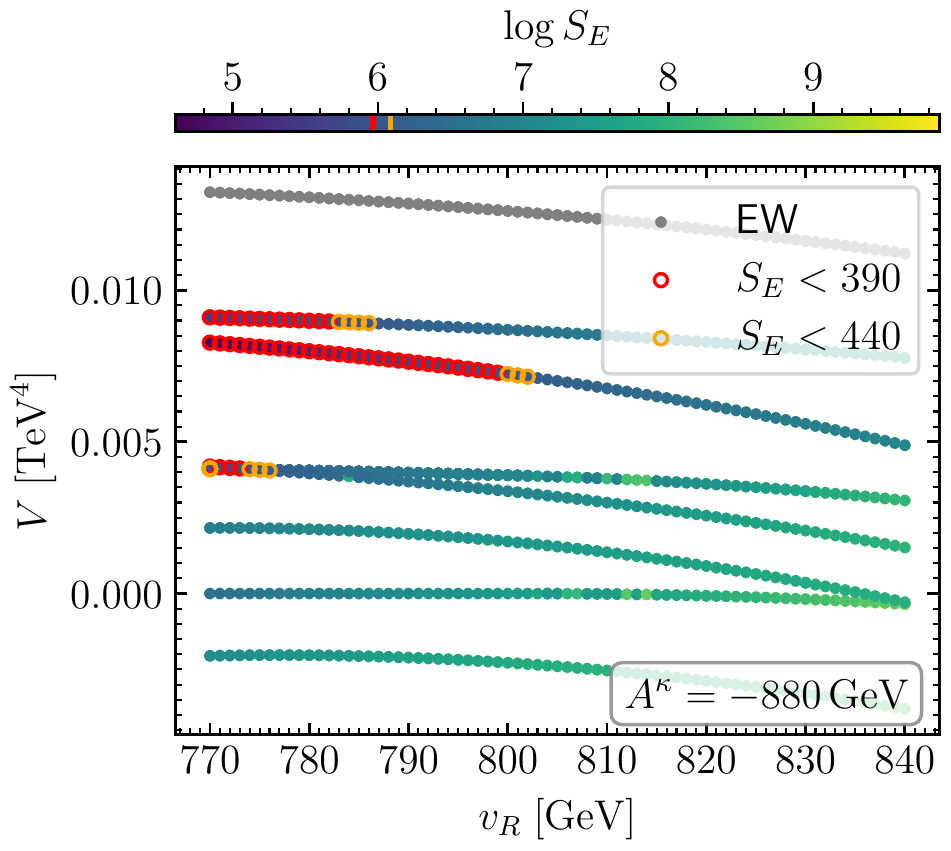} \\[0.4em]
\includegraphics[height=7.2cm]{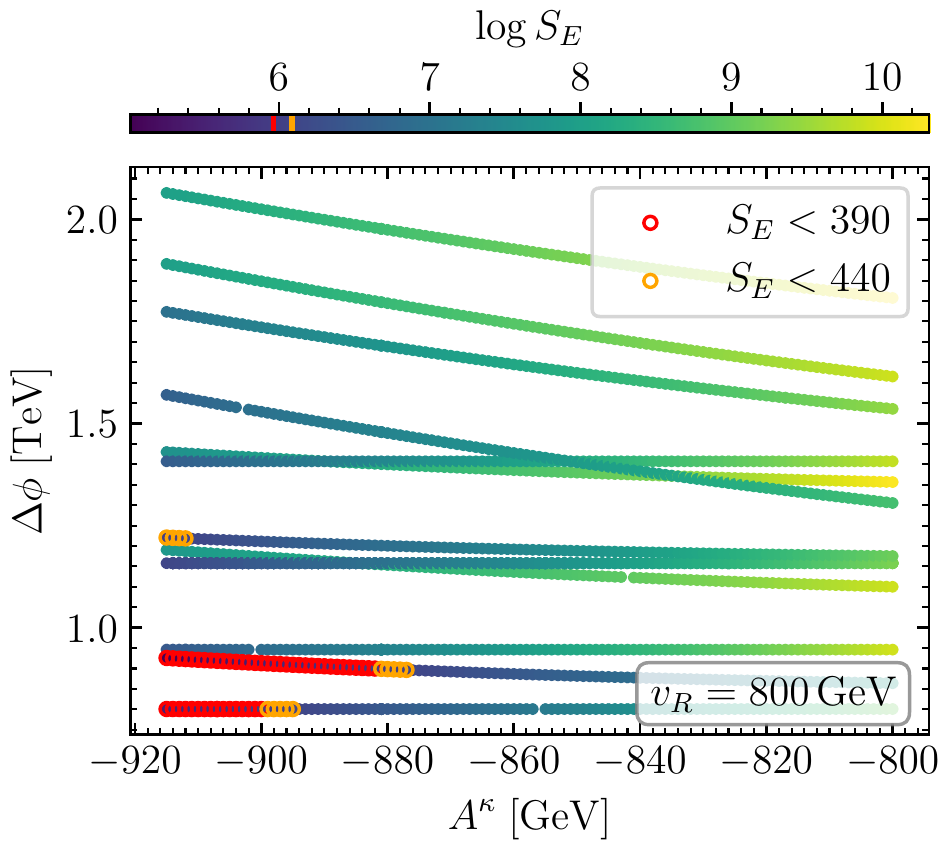}~
\includegraphics[height=7.2cm]{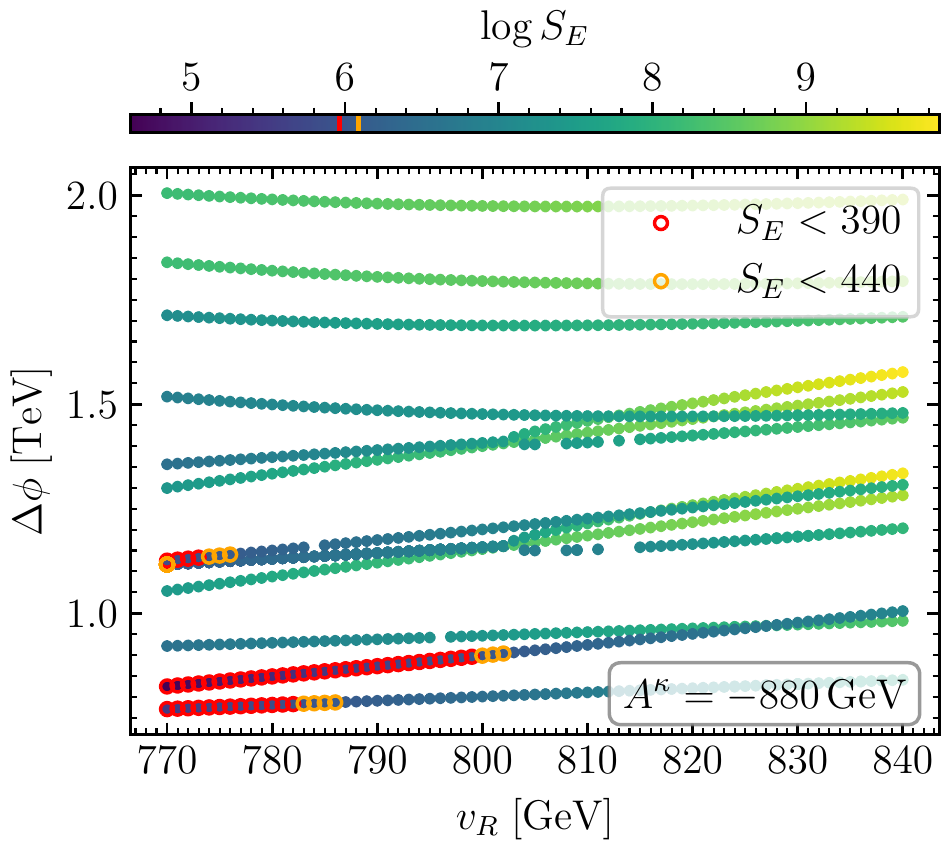}
\caption{\small EW minimum and deeper unphysical minima
as a function of $A^\kappa$ for fixed $v_{R}$
(left) and as a function of $v_{R}$ for fixed
$A^\kappa$ (right). Shown are for each local minimum
the potential values $V$ (top) and the field space distance
between the unphysical minimum and
the EW minimum $\Delta \phi$ (bottom). The colors of the
points indicate the value of the euclidean bounce action
$S_E$. Points with $S_E < 390$ are highlighted with
a red edge, while points with $390 \leq S_E \leq 440$
are highlighted with an orange edge.}
\label{fig:vacst}
\end{figure}

In \reffi{fig:vacst} we show the results
of the analysis of vacuum stability
for the parameter space under investigation.
On the left-hand
side the points from the $A^\kappa$--scan
are shown, and on the right hand side we
show the points from the scan over $v_R$.
In the upper row we show
the values of the potential $V$
for the EW minimum and 
for all unphysical minima with potential values
below the EW minimum. 
The points corresponding to the EW vacuum
are shown in gray.
For the unphysical
minima the colors of the points indicate
the values of the euclidean bounce action
$S_E$ for the 
tunneling from the
EW minimum into the unphysical minima.
Parameter points for which we find that the 
EW vacuum is short-lived compared to the lifetime of the universe, 
corresponding to $S_E < 390$, are furthermore highlighted in red.
The regions with $390 < S_E < 440$, for which
in our approach no clear 
distinction between a short-lived and
a sufficiently long-lived vacuum 
can be drawn, are highlighted in orange
(see the discussion in \refse{bouncer}).

In both scans it can be observed that
the EW minimum is not the global
minimum of the potential
for any of the points.
In the $A^\kappa$--scan we find a total
number of 12 deeper unphysical minima, 
where a few of those have the same depth
$V$ due to the accidental symmetries
mentioned in \refse{intromunu}, such
that only eight lines are visible
in the upper left plot of \reffi{fig:vacst}.
The situation is similar for
the upper right plot, 
with the exception that for this scan
the lines of two unphysical minima merge for small 
values of $v_R$. 
The different unphysical minima below the EW one can be 
more easily distinguished from each other
in the plots displayed in the
lower row where the field space distance
between the unphysical minimum and
the EW minimum, $\Delta \phi$, is shown.
Here one can see for the $A^\kappa$--scan the already mentioned 
number of 12 unphysical minima below
the EW one, while for the $v_R$--scan
up to 14 unphysical minima below the EW one are visible. 
The lower right plot as a function
of $v_R$ shows the interesting feature that 
for a certain value of $v_R \approx 803\gev$
two of the unphysical minima bifurcate into
two different ones each. Hence, 
in the $v_R$--scan either 12 
or 14 unphysical minima below the
EW minimum are present (see also the discussion
below). 

Even though there are numerous
unphysical minima present
below the EW vacuum for all the investigated
parameter points, most of the
unphysical minima do not 
give rise to a rapid decay
of the EW vacuum,
because the tunneling rates for all
possible transitions turn out to be very small.
This is in particular the case also
for the global minimum of each parameter
point, which is characterized by non-zero
values of the right-handed sneutrino fields
$\phi_{iR}$
and large negative values of the Higgs
fields $\phi_d$ and $\phi_u$. As a result
of the latter, the tunneling rates for
transitions into the
global minimum are suppressed by the large
distance in field space between the
EW minimum with positive values of
$\phi_d$ and $\phi_u$.

The displayed results clearly indicate that
a detailed investigation of the 
tunneling rates is crucial for
determining the constraints from vacuum stability 
on the parameter space of the $\mu\nu$SSM. 
Analyses in which just the minima of
the potential are determined and parameter 
points where the EW vacuum is not the
global one are discarded, as they have sometimes 
been carried out in the literature for
other models, would obviously be completely 
misleading for the case of the $\mu\nu$SSM.

As another interesting feature that
is visible in \reffi{fig:vacst}, 
one can observe
in the scan over $A^\kappa$
the presence of a local minimum with
a vanishing potential value $V=0$
over the whole scan range. This minimum
is located at the origin of field space,
and it is therefore
of particular interest.\footnote{This
feature does not occur in the MSSM, where
the origin is always a saddle point.}
According to our approach
for investigating vacuum stability,
we find that the presence of this minimum at the origin of
field space does not give rise to
exclusions of parameter points
in this scan, because
the tunneling rate from the EW vacuum
into this minimum is too small.
It should however be noted that the
analysis carried out here relies on the assumption
that the EW minimum was actually adopted
by the universe in its thermal history.
If a local minimum at the origin exists
in the universe now, 
one may wonder, however, whether such
a local minimum at the origin may 
have existed already during the
thermal history of the early universe.
While in our present analysis of
vacuum stability we restrict ourselves to the 
case of temperature $T = 0$, 
a detailed 
analysis of the thermal history of
the potential would be required 
in order to determine whether the EW
vacuum was actually reached during the thermal 
evolution.
Consequently, an analysis
of the evolution of the field values
as a function of the temperature (which
is inversely proportional to the time)
could potentially give rise to even
stronger exclusion bounds compared to the
zero-temperature analysis performed here
(see also the discussion in
\refse{conclu}).
The minimum in the origin is also present
for parts of the parameter space of the
scan over $v_R$. It exists in the
range $770\gev \leq v_R \leq 802\gev$,
as can be 
seen in the upper right plot of \reffi{fig:vacst}.
For larger values of $v_R$
the potential values bend down from 
the horizontal line at $V=0$
towards negative potential values,
indicating that the minimum shifts
away from the origin. The
origin becomes unstable along the
direction of $\phi_u$. Accordingly,
in the range $803\gev \leq v_R \leq
840\gev$ we find
an unphysical minimum with non-zero
values of $53\gev \leq |\phi_u| \leq
377\gev$ and
all other fields vanishing.

The above discussion leads to the following
important conclusions. We have found
that the $\mu\nu$SSM gives rise to a 
rich vacuum structure, featuring a metastable (i.e., long-lived 
in comparison to the age of the universe)
or short-lived 
EW vacuum together with
several deeper unphysical minima.
Thus, the analysis of the
validity of a certain parameter point should
include a test of vacuum stability.
Furthermore, as mentioned above,
it would not be sufficient to simply require
that the EW minimum is the global minimum
of the potential, as such a criterion would
exclude large parts of actually allowed
parameter space.
Instead, a calculation of the tunneling rates
of the EW vacuum has to be performed in order to
reliably determine whether a parameter point 
is phenomenologically viable.
In the following we further discuss the structure
of the potentially dangerous unphysical minima,
including an analytical discussion of the
dependence on
the main parameters entering the calculation
of $S_E$, in order to determine which parts
of the $\mu\nu$SSM parameter space are expected
to be susceptible to vacuum instabilities.

We note in this context that parameter
points where the EW vacuum
is short-lived feature
transitions into unphysical
minima which are not the deepest ones. On the
contrary, we can see from the plots in the
upper row of \reffi{fig:vacst}
that the most dangerous minima
are in fact the ones with the highest values of $V$, 
which therefore have
the smallest potential difference compared
to the EW minimum, $\Delta V_{\rm EW}$.
This indicates 
that for the considered case $\Delta V_{\rm EW}$ is not the
most important quantity in the evaluation of the
transition rate. Instead, as
can be seen in the lower row of \reffi{fig:vacst},
the distance in field
space $\Delta \phi = |\phi_{\rm EW} - \phi|$ between
the EW minimum at $\phi_{\rm EW}$ and an unphysical
minimum at $\phi$
has a bigger impact. In the lower row
of \reffi{fig:vacst}, $\Delta \phi$
is shown as a function of $A^\kappa$ on the left
and of $v_R$ on the right for all potentially
dangerous (i.e., deeper than the EW vacuum) unphysical minima
that occur for each parameter point. It can be seen that
the transitions with the lowest values of $S_E$ 
are typically those into 
the unphysical
minima with the lowest values of $\Delta \phi$
(as expected, a smaller distance in field
space is correlated with a higher
decay rate of the EW vacuum). 
Our results indicate that for the
analyzed $\mu\nu$SSM scenarios
the distance in field space is more important 
for the determination of the tunneling rate of the EW vacuum 
than the difference of the
potential depths. 

In the lower right plot of \reffi{fig:vacst}, 
and to a lesser extent also 
in the lower left plot,
one can see that a few points are missing in the curves (in particular 
near the bifurcation point in the lower right plot). 
This is due to the fact that the
numerical
approach applied here to determine
the stationary conditions
missed one of the minima (see also
the discussion in \refse{minimization}).
Nevertheless, given the fact that this happened
only in rare cases, our conclusions regarding
the impact of vacuum stability constraints on
the $\mu\nu$SSM are not affected by this issue.
In practice, it was always possible to find also
the minima missing in \reffi{fig:vacst} by marginally
changing the input parameters. However, we decided
to show the results as they were obtained during
a single run in order to give an impression of the
performance of the applied procedure.

\begin{table}
\centering
{\renewcommand{\arraystretch}{1.4}
\footnotesize
\begin{tabular}{>{\bfseries}l||cc|cccccccc}
 & $A^\kappa$ & $v_{R}$ & $\phi_d$ & $\phi_u$ &
  $\phi_{1R}$ & $\phi_{2R}$ & $\phi_{3R}$ & $\phi_{1L} / 10^{-4}$ &
    $\phi_{2L} / 10^{-4}$ & $\phi_{3L} / 10^{-4}$ \\
\hline
\hline
AI & $-880$ & $800$ & $115.3$ & $230.9$ & $803.7$ &
  $801.1$ & $-0.4755$ & $1.053$ & $1.050$ &
    $-0.000642$ \\
AII &  &  & $156.2$ & $515.7$ & $-25.18$ & $992.0$ & $804.8$ &
  $-0.01805$ & $2.259$ & $2.167$ \\
\hline
BI &$-900$ & $800$ & 0 & 0 & 0 & 0 & $800.0$ & 0 & 0 & 0 \\
BII &  &  & $162.8$ & $530.0$ & $-25.65$ & $1014$ & $-0.2768$ &
  $-0.01578$ & $2.718$ & $-0.00027$ \\
BIII &  &  & $165.6$ & $534.3$ & $-27.45$ & $1017$ & $806.6$ &
  $-0.0202$ & $2.752$ & $2.263$ \\
BIV &  &  & $115.5$ & $230.9$ & $806.1$ & $800.7$ & $-0.4541$ &
  $1.057$ & $1.050$ & $-0.00061$ \\
\hline
CI & $-880$ & $780$ & 0 & 0 & 0 & 0 & $779.9$ & 0 & 0 & 0 \\
CII &  &  & $115.5$ & $230.6$ & $788.9$ & $781.0$ &
  $-0.4449$ & $1.059$ & $1.0502$ & $-0.00061$ \\
CIII &  &  & $135.6$ & $449.8$ & $-13.08$ & $958.5$ &
  $-0.159$ & $0.00472$ & $2.215$ & $-0.00006$ \\
CIV &  &  & $138.9$ & $455.6$ & $-14.48$ & $961.9$ & $785.2$ & $0.00235$ & $2.256$ & $1.902$ \\
\hline
\hline
EW &  & $v_R$ & $110.1$ & $220.2$ & $v_R$ & $v_R$ &
  $v_R$ & $1$ & $1$ & $1$
\end{tabular}
}
\caption{\small Field values of unphysical
local minima for which
$S_E < 700$ are given for three selected
parameter points.
The last row shows the field values of
the EW minimum, 
which are the same for the three
parameter points.
Dimensionful parameters and field
values are given in GeV.}
\label{tab:unphy}
\end{table}

The fact that the field space distance $\Delta \phi$
appears to be more important than the potential
difference $\Delta V_{\rm EW}$, and also that
the dangerous minima were found at distances
$\Delta \phi \approx v_R$, makes it apparent
that 
the field values of the
right-handed sneutrino fields are particularly
important 
for the determination of the tunneling rates. In order
to shed more light on the field
configurations of the
unphysical minima, we
show in \refta{tab:unphy} the field values
of the most dangerous minima for three selected
parameter points 
together with the field values of the EW minimum
($\phi_d$ and $\phi_u$ in the EW minimum
are determined by $\tan\beta = 2$
and $v \approx 246\gev$).
We only display the unphysical minima for which
the transition rate
from the EW minimum corresponds
to values of $S_E$ below $700$.
The first parameter point under consideration (A)
has $A^\kappa = -880\gev$ and $v_R = 800\gev$,
and it features
the unphysical minima AI and AII
with $S_E < 700$.
The most striking feature of these minima is
that, while the EW minimum has universal field
values $\phi_{iR} = v_R$, in AI and AII the field
value of one of the right-handed sneutrino fields
is almost zero, while the other two approximately
maintain values of about $v_R$.
A similar observation can be made for the second
parameter
point (B), with $A^\kappa = -900\gev$ and $v_R
= 800\gev$,
featuring the unphysical minima BI--BIV.
Here, the minima BIII and BIV are the analogues
to the minima AII and AI
of the first parameter point, respectively,
i.e.~they lie on the same branch of points
in the plots in \reffi{fig:vacst}.
In addition, two more minima are shown
for the parameter point B, where
only one of the fields $\phi_{iR}$ has a
value of $\approx v_R$, while the other
two have $\phi_{iR} \approx 0$.
For example,
BI lies exactly on the $\phi_{3R}$ axis.
Two minima of this kind are also present
for the parameter point A, but they
correspond to EW vacuum decay rates
with $S_E > 700$ and are therefore not
shown in the table. The second point has a larger
value of $|A^\kappa|$. This suggests that
more negative values of $A^\kappa$ can
give rise to an unstable EW vacuum (as also shown
in \reffi{fig:vacst}). Finally, the third
point (C) with $A^\kappa = -880\gev$ and
$v_R = 800\gev$ features the unphysical
minima CI--CIV. While they show a
similar field configuration as BI--BIV,
it is interesting to
compare the different parameters related to
the tunneling probabilities.

\begin{table}
\centering
{\renewcommand{\arraystretch}{1.4}
\footnotesize
\begin{tabular}{>{\bfseries}l||ccccc|c}
 & $\Delta \phi$ & $m(\hat{\varphi})$ &
   $A(\hat{\varphi})$ & $\lambda(\hat{\varphi})$ &
     $-\Delta V_{\mathrm{EW}}^{1/4}$ & $S_E$ \\
\hline
\hline
AI & $800.6$ & $106.2$ & $58.71$ & $0.0462$ & $250.2$ &
  $660.0$ \\
AII & $898.4$ & $83.81$ & $45.28$ & $0.03345$ & $270.8$ &
  $\mathbf{401.5}$ \\
\hline
BI & $1158$ & $91.35$ & $39.44$ & $0.02244$ & $313.9$ &
  $675.6$ \\
BII & $1211$ & $81.73$ & $34.37$ & $0.01901$ & $319.1$ &
  $609.2$ \\
BIII & $912.9$ & $70.21$ & $42.26$ & $0.0318$ & $278.1$ &
  $\mathbf{241.1}$ \\
BIV & $800.6$ & $94.07$ & $56.68$ & $0.04620$ & $258.1$ &
  $\mathbf{380.4}$ \\
\hline
CI & $1130$ & $86.97$ & $38.34$ & $0.02248$ & $308.1$ &
  $610.9$ \\
CII & $780.6$ & $90.08$ & $55.01$ & $0.04619$ &
  $252.6$ & $\mathbf{353.89}$ \\
CIII & $1150$ & $77.41$ & $34.34$ & $0.02013$ &
  $308.8$ & $516.1$ \\
CIV & $848.8$ & $64.43$ & $42.97$ & $0.0351$ &
  $266.9$ & $\mathbf{180.7}$
\end{tabular}
}
\caption{\small Values for the coefficients
$m$, $A$ and $\lambda$ as defined in
\refeq{eq1dpot} for the non-EW minima
shown in \refta{tab:unphy}. Also shown are
the distance in field space
$\Delta \phi$ and the potential
difference $-\Delta V_{\rm EW}^{1/4}$
between the EW minimum
and the unphysical minima.
The last column shows the euclidean bounce action
$S_E$ for the transitions.
Values of $S_E$ giving rise
to a (potentially) short-lived EW vacuum are
given in bold font. 
Dimensionful parameters and field
values are given in GeV.}
\label{tab:unphyparas}
\end{table}

For this reason we show in \refta{tab:unphyparas}
the values of $S_E$ associated to each minimum,
together with the coefficients that define
the form of $V_{\rm EW}$ as defined in
\refeq{eq1dpot}. Values of $S_E$ giving rise
to a (potentially) short-lived EW vacuum are
highlighted in bold font. In addition, the table also
lists the values for the field space distance
$\Delta \phi$ and the potential difference
$-\Delta V_{\rm EW}$ between the EW minimum and the
unphysical minima. One can see that all three
points have at least one unphysical deeper
minimum associated
with a value of $S_E < 440$. For AII we find
$S_E = 401.5$, such that 
for the parameter point A
the EW vacuum
is potentially short-lived. In order to definitively
answer the question whether point A features
a sufficiently long-lived EW vacuum or not,
one would have to improve the computation
of $S_E$, 
incorporating in particular a more elaborate 
treatment of
the prefactor $K$
introduced in \refeq{eq:decayrate}
(see also the discussion below).
For the parameter point B,
the presence of both BIII and BIV with
$S_E = 241.1$ and $S_E = 380.4$, respectively,
indicates that the EW vacuum is short-lived,
and the parameter point should be rejected.
The same observation can be made for the
parameter point C, for which we find
$S_E = 353.89$ for CII and $S_E = 180.7$
for CIV.

One can gain further analytical insight
into the 
tunneling dynamics by
comparing the minima of different
parameter points with similar
field configurations, i.e.~minima
that lie on the same branch of
points in \reffi{fig:vacst}.
As already mentioned before,
for the point A with the
minimum AI and $S_E = 660.0$
the analogues minimum for the point B
is the minimum BIV with $S_E = 380.4$.
Hence, the change from $A^\kappa = -880\gev$
to $A^\kappa = -900\gev$ 
gives rise to an
instability of the EW vacuum.
The reason for this can be understood
analytically by realizing that the only
field that substantially changes 
if the EW vacuum decays into AI and BIV 
is $\phi_{3R}$. In the
EW minium we have $\phi_{3R} = v_R = 800\gev$,
and the field evolves to $\phi_{3R} \approx 0$
during the transition.
Thus, the potential $V_{\rm EW}$ as defined
in \refeq{eq1dpot} is given for this
kind of transitions
in very good approximation by choosing
the unit field vector $\hat{\varphi}_i$ in the
direction of $-\phi_{3R}$, such that
$\hat{\varphi}_{\rm AI,BIV} = (0,0,0,0,-1,0,0,0)$.
Then the coefficients of $\Delta V_{\rm EW}$
are given by
\begin{align}
A(\hat{\varphi}_{\rm AI,BIV}) &= \kappa_3 \left(
\frac{1}{3 \sqrt{2}}A^\kappa_3 +
\kappa_3 v_{3R} \right) \ , \label{eqA3r} \\
m^2(\hat{\varphi}_{\rm AI,BIV}) &= \frac{\lambda_3 v^2}{8 v_{3R}}
\left(
\sqrt{2} A^\lambda_3 s_{2\beta} -
2\left(
\lambda_1 v_{1R} + \lambda_2 v_{2R}
\right)
\right) +
\frac{1}{4} \kappa_3 v_{3R}
\left(
\sqrt{2} A^\kappa_3 +
4 \kappa_3 v_{3R}
\right) \ , \label{eqmsqm3r}\\
\lambda(\hat{\varphi}_{\rm AI,BIV}) &= 
\frac{1}{4} \kappa_3^2 \ ,
\end{align}
where in the second row the terms $\sim v_{iL}$
and $\sim Y^\nu_{ij}$ were neglected,
and where in our scans $v_R = v_{1R}
= v_{2R} = v_{3R}$.
Furthermore, given the small value of
$\lambda_{3} = 0.05$ in this scenario,
the first term in the expression for
$m^2$ is suppressed compared
to the second term.
For the determination
of $S_E$
the ratio $A^2 / (m^2 \lambda)$ is particularly important
(see also the definition
of $\delta$ in 
\refeq{eq:SE}),
which is related to
the fact that terms
in $V_{\rm EW}$ with odd powers of the fields
give rise to the potential barrier
that separates the EW minimum from
the unphysical minimum: increasing
the trilinear coefficient
$A$ leads to larger values of $S_E$,
while $S_E$ becomes smaller with
increasing coefficients $m^2$
and $\lambda$ of the terms with even
powers of the fields. Focusing on the
terms $\sim A^\kappa_3$, we find that this
ratio scales with $-\kappa_3 |A^\kappa_3|
/ v_{R}$.\footnote{Note that $A^\kappa_i < 0$
is required in order to avoid tachyonic CP-odd states.}
Thus, if $|A^\kappa_3|$ becomes larger, the
decrease in the potential barrier 
caused by $A$ is more important than the decrease
of the bilinear coefficient $m^2$, such that
$S_E$, and therefore the lifetime
of the EW vacuum, becomes smaller.

It should be noted that the change from
an unstable to a metastable EW
vacuum sensitively depends
on the value of $A^\kappa$,
as we demonstrated here for the points A
and B, where the $A^\kappa$ values differ
by only $\approx 2\%$ 
for points with a metastable
and an unstable vacuum.
Consequently, one can expect that the
uncertainty 
that is related to our treatment of the
prefactor $K$ in \refeq{eq:decayrate}, giving rise
to the intermediate regime $390 < S_E < 440$
in which the lifetime of the EW vacuum is
of comparable size as the age of
the universe,
affects only a relatively small fraction
of the $\mu\nu$SSM parameter space.
For most parts of the parameter space,
the
value of $S_E$ for possible vacuum decays
should either be substantially below or
above the intermediate regime with $S_E \approx 400$, 
such that its impact on vacuum stability
can clearly be determined.\footnote{A
similar observation was made
in \citere{Hollik:2018wrr}
regarding charge- and color-breaking minima in the MSSM and their
relation to the precise values of the
soft trilinear couplings.}

The smallest values of $S_E$ are found for
the parameter point C, which has a smaller value
of $v_R = 780\gev$ compared to $v_R =
800\gev$ for the points A and B.
We already mentioned before that this can
intuitively be understood due to the fact
that the field space difference $\Delta \phi$
becomes smaller for the relevant unphysical
minima. We find, for example, a value of
$S_E = 353.89$ for the minimum CII, while for the
analogous minimum AII of the first parameter
point with an identical value of $A^\kappa$
we find $S_E = 401.5$.
Following the above reasoning, the smaller $S_E$
value for CII compared to AII can be understood
by the fact that (for $\lambda_3 \ll 1$)
$A^2 / (m^2 \lambda)$ scales with
$1 / v_{R}$. 
Comparing to \refta{tab:unphyparas}, one can
additionally see that also the absolute
value of the potential difference $|\Delta 
V_{\rm EW}|$ changes with $v_R$.
However, since the 
rate for the vacuum decay into CII is larger
compared to the decay into AII, even though
$|\Delta V_{\rm EW}|$ is larger for AII,
one can conclude (as before) that the
field space difference $\Delta \phi$
has quantitatively
a larger impact on the lifetime of the
EW vacuum.

\begin{figure}
\centering
\includegraphics[width=0.48\textwidth]{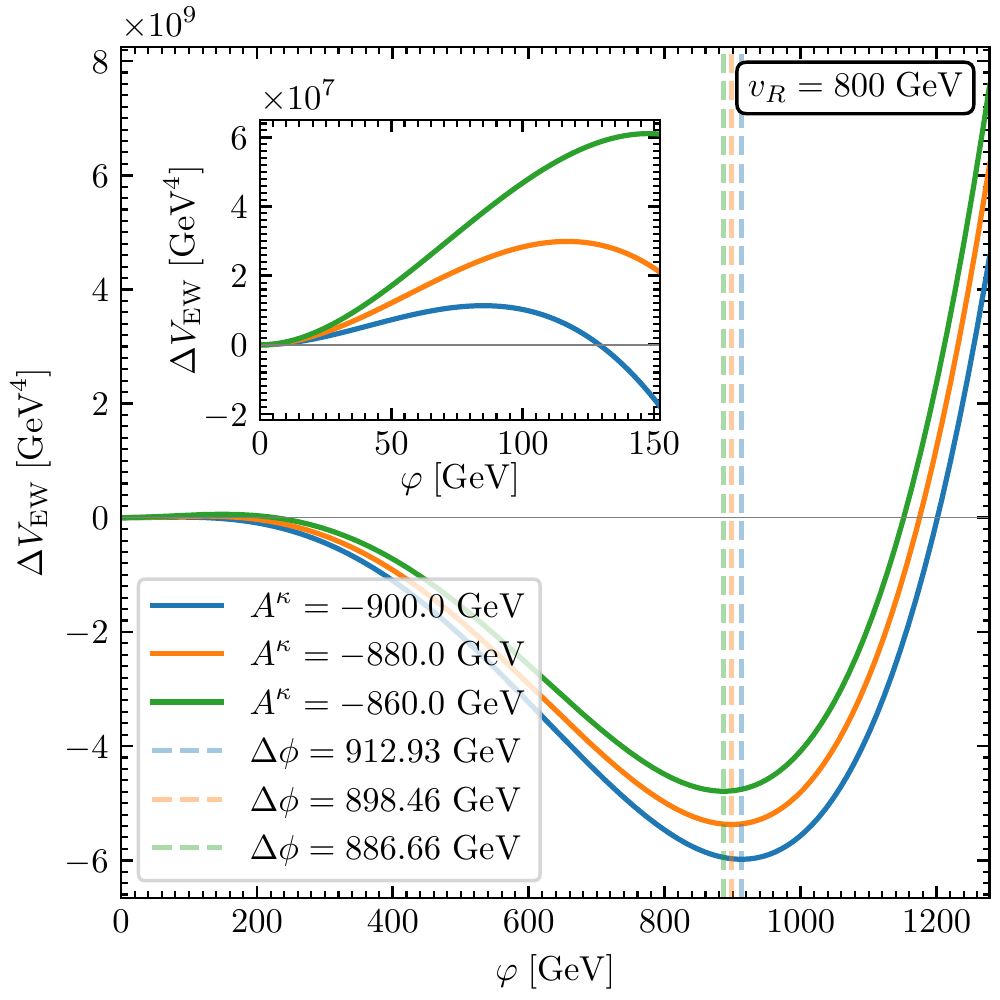}~
\includegraphics[width=0.48\textwidth]{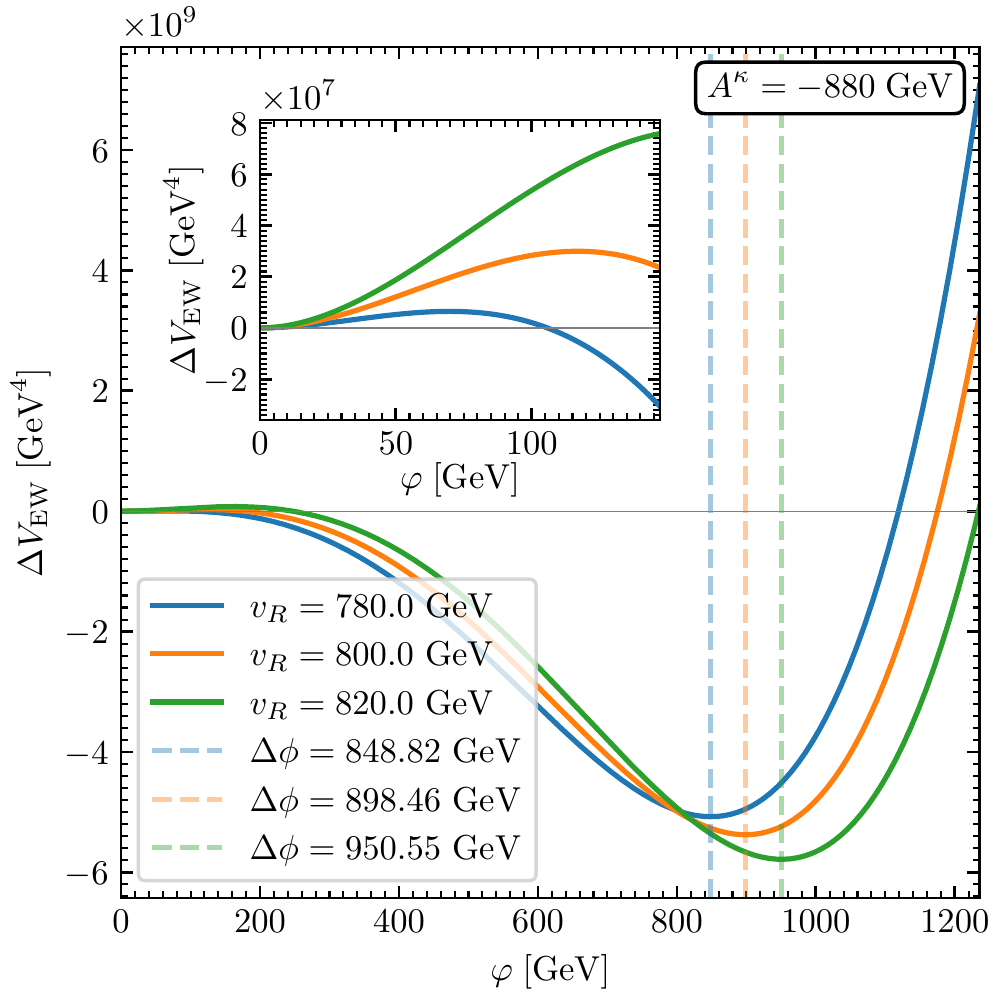}
\caption{\small $\Delta V_{\rm EW}$ along the straight path
connecting the EW minimum and the most dangerous unphysical
minimum for different values of $A^\kappa_i$ (left)
and $v_{iR}$ (right). Also indicated are the field
space distances $\Delta \phi$ with
vertical dashed lines.
}
\label{fig:potew}
\end{figure}

The most dangerous minima of all three
parameter points cause EW
vacuum decays for which
$\phi_{1R}$ (instead of
$\phi_{3R}$) evolves from $\phi_{1R} = v_R$
to $\phi_{1R} \approx 0$.
For these transitions an important
difference arises from the fact that
$\lambda_1 = 0.56 \gg \lambda_3$, such that
the interactions between $\phi_1$ and
the Higgs doublet fields $\phi_d$ and $\phi_u$
cannot be neglected.
As a result, also
the latter fields change during the
transition from the EW vacuum into
the vacua AII, BIII and CIV, while $\phi_d$
and $\phi_u$ remain approximately constant
for the transitions into AI, BIV and CII.
For the cases where
the doublet fields change
during the transition, the expressions for
the coefficients of $V_{\rm EW}$ become significantly
more complicated and the analytic treatment
does not provide much insight.\footnote{We
nevertheless provide in
\refap{app:coefs} the coefficients of $V_{\rm EW}$
for 
example field directions.} Instead we show
in \reffi{fig:potew} the potential difference $\Delta V_{\rm EW}$
in the direction of the unphysical minimum
corresponding to the smallest value of $S_E$
for different values of $A^\kappa$ and $v_R$.

In the left plot of \reffi{fig:potew} we show the potential
difference $\Delta V_{\rm EW}$ for three
different values of ${A^\kappa = -900,-880,-860\gev}$
and for fixed $v_R = 800\gev$, while the remaining
parameters are fixed as shown in
\refta{tab:bp2pa}. Hence, the curve
with $A^\kappa = -880\gev$ (orange) corresponds
to the parameter point A discussed in relation
to \refta{tab:unphy} and \refta{tab:unphyparas},
and $\Delta V_{\rm EW}$ is
shown in the direction of the minimum AII.
One can see that the variation of $A^\kappa$
induces a variation of both the barrier height
and the depth of the unphysical minimum, whereas
the field space difference $\Delta \phi$
(indicated by the vertical dashed lines) is
largely unaffected.
This follows the expectation from
\refeq{eqA3r}: For a transition which at least
partially evolves in the
direction~${-\phi_{iR} \ ,
\ i = 1,2,3}$, there is
a positive contribution to the
coefficient $A$ which is
proportional to $v_{iR} \kappa_i^2$, and a negative
contribution proportional to $-\kappa_i |A^\kappa_i|$.
Thus, the smaller $|A^\kappa|$,
the larger is the potential barrier between both
minima for fixed values of $v_{iR}$ and $\kappa_i$,
and the lifetime of the EW vacuum increases.
On the other hand, for large values
of $|A^\kappa_i|$ the lifetime of the
EW vacuum decreases,
and constraints arising from the
vacuum stability become important.

This observation is not surprising since it is known
that for $|A^\kappa_i| \gg v_{iR} \kappa_i$
the CP-even right-handed
sneutrinos can become tachyonic~\cite{Ghosh:2014ida,
Biekotter:2019gtq},
pointing to the fact that vacuum instabilities
might occur. Similar observations were also
made in the NMSSM where only one gauge singlet
field is present~\cite{Ellwanger:1996gw,
Baum:2020vfl}.\footnote{The heuristic NMSSM
criterion $A_\kappa^2 >
9 m_S^2$~\cite{Ellwanger:1996gw}, with
$m_S^2$ being the soft mass
parameter of the
singlet scalar, relies on the condition that the
EW minimum is the global minimum of the
potential, and does not take into account
the possibility of a
metastable EW vacuum.}
In the $\mu\nu$SSM,
the presence of three such fields leads to
the fact that there are more ways in which
dangerous unphysical minima can arise.
For instance, there can be minima in which
one, two or all three fields $\phi_{iR}$
take on values of $\approx 0$. In addition,
the way in which the singlet states are
coupled to the doublet Higgs fields yields
a much richer 
structure of vacuum
configurations, with several options that
could give rise to EW vacuum
instabilities. Thus, compared
to the NMSSM, the constraints from vacuum
instabilities can be expected to be even more
important in the $\mu\nu$SSM.

In the right plot of \reffi{fig:potew} we show
$\Delta V_{\rm EW}$ in the direction of the
most dangerous unphysical minimum
for
different values of ${v_R = 780,800,820\gev}$
and a fixed value of $A^\kappa = -880\gev$.
As before, the orange curve belongs
to the parameter point A of \refta{tab:unphy},
with $\Delta V_{\rm EW}$ shown
in the direction of the minimum AII.
There are two main effects of the variation
of $v_R$ on the stability of the EW vacuum.
As already mentioned before, smaller values
of $v_R$ reduce the value of $\Delta \phi$
(indicated by the vertical dashed lines
in \reffi{fig:potew})
for minima of the type shown in \refta{tab:unphy},
in which the field values of
one or more right-handed sneutrinos
go from $v_R$ to
zero during the transition.
Hence, smaller values
of $v_R$ are also associated
with a smaller lifetime of the (metastable)
EW vacuum. This effect is further enhanced
by the fact that also the potential barrier
becomes smaller when the value of $v_R$
becomes smaller.

In combination of both observations
discussed above,
i.e., smaller values of $v_{iR}$ and larger
values of $|A^\kappa_i|$ can lead to an unstable
EW vacuum, there is a clear phenomenological
consequence. One can expect that 
an instability of 
the EW vacuum in the $\mu\nu$SSM
can be caused by the presence
of relatively light
right-handed sneutrinos, with masses at
or below the EW scale. Since this is the
parameter region in which there are prospects
to potentially probe the existence
of these particle
states at the LHC or other future colliders,
either directly or indirectly via
signal rate measurements of the
Higgs boson at $125\gev$,
the constraints from vacuum stability can 
have an important impact on
scenarios that can be probed by collider
experiments, and thus should be taken
into account in analyses of the
collider prospects of the $\mu\nu$SSM.

\begin{figure}
\centering
\includegraphics[width=0.8\textwidth]{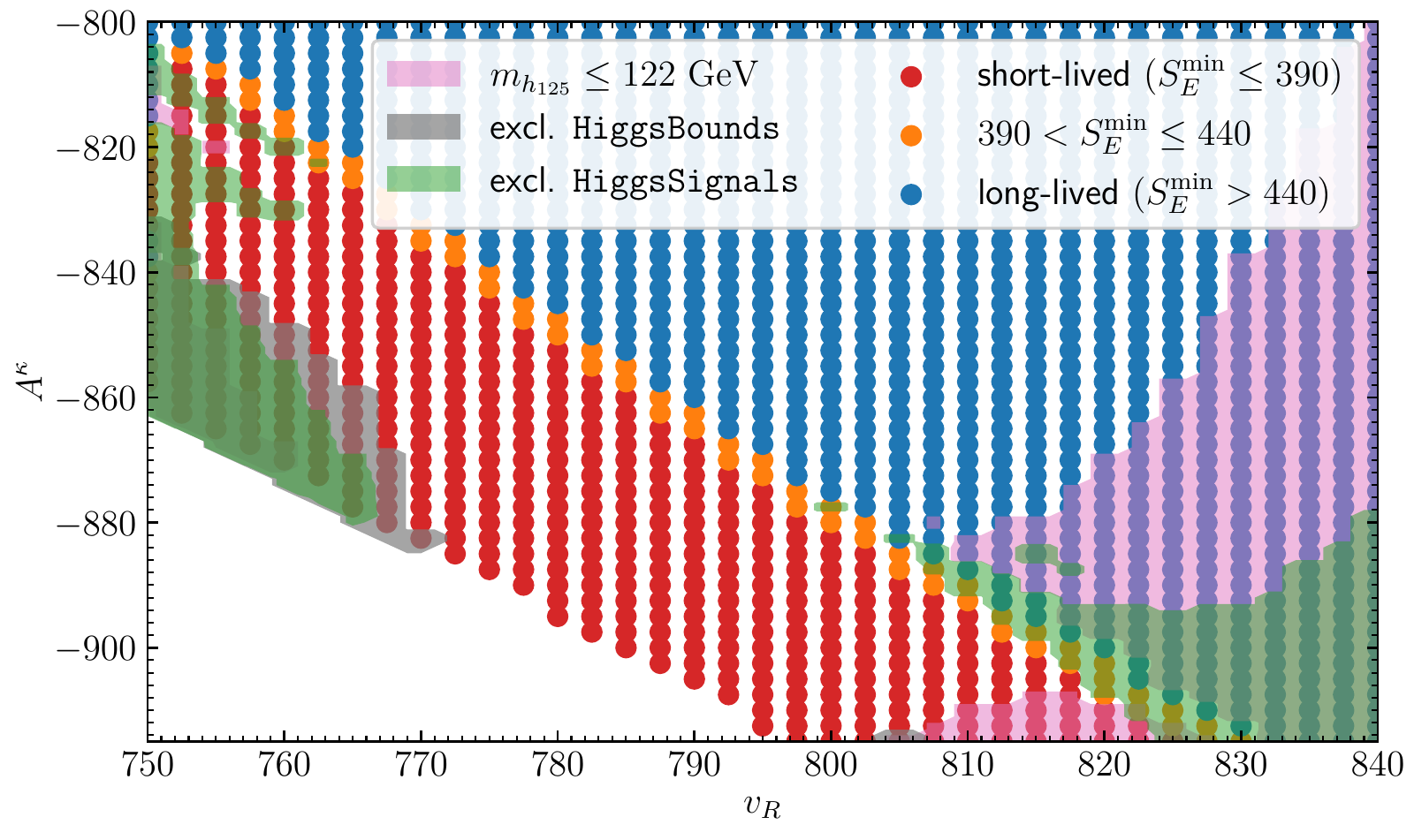}
\caption{\small Vacuum stability constraints
in the
$v_R$-$A^\kappa$ plane. Blue points feature a
sufficiently long-lived EW vacuum, while the red
points give rise to a rapidly decaying EW vacuum.
The orange points feature a potentially short-lived
EW vacuum (see text). The gray and green
areas indicate excluded
regions based on the \texttt{HiggsBounds}
and \texttt{HiggsSignals} constraints (see
text for details), and the pink area
indicates the parameter region in which
$m_{h_{125}} < 122 \gev$.}
\label{fig:phase}
\end{figure}

As an illustration of our results,
we show
in \reffi{fig:phase}
the parameter points
in the $v_R$-$A^\kappa$ plane, in which
the parameter space is divided into
a red region in which the EW vacuum
is short-lived, an orange region in
which the EW vacuum is potentially
short-lived
(see the discussion
in \refse{bouncer} about this intermediate region),
and a blue region in
which the EW vacuum is metastable
and long-lived.
There are no points in the lower
left corner due to the appearance
of tachyonic CP-even states at
tree level (see also
\reffi{fig:spec}).
For none of the points the
EW minimum is the global minimum
of the potential.
Demanding that the EW minimum should be
the global minimum would 
imply the exclusion of the 
phenomenologically viable
blue region.
This illustrates the
importance of taking into account
the possibility of a sufficiently long-lived
metastable EW vacuum.
Since the lifetime of the
EW vacuum has a sensitive dependence on the
precise values of $v_R$ and $A^\kappa$,
the orange
region constitutes only a small narrow
band of the investigated parameter space.
As the parameter points in the
red region feature a short-lived EW vacuum,
they are not physically viable and should
be excluded.

The lightest (loop-corrected)
scalar masses that we find in the red, blue
and orange regions  are $m_{h_1} = 70.77 \gev,
112.4 \gev$ and $116.5 \gev$, respectively.
As a result, we find that in this scenario
all points with $m_{h_1} < 112 \gev$
are excluded due to EW vacuum stability
constraints.
This demonstrates the importance of
the interplay
between the constraints from vacuum stability
and possible collider phenomenology of
the model at current or future colliders.
As is indicated by the gray area, only
two small parameter regions are excluded via
cross section limits from direct searches,
and both of these regions lie within the
red parameter region. Hence, we observe in
the scenario investigated here that the
vacuum stability constraints have a larger
impact and 
give rise to
exclusion limits
that by far exceed the present limits from direct
searches for the BSM Higgs bosons of
the $\mu\nu$SSM.

In \reffi{fig:phase} we also indicate
the parameter points for which the prediction
for the mass of the SM-like Higgs-boson does not
lie within the range $(125 \pm 3) \gev$
(pink area).
However, the points in this region
should not be considered as
strictly excluded, since $m_{h_{125}}$ could
easily be adjusted to lie within the
required interval by modifying the parameters
related to the stop sector, from which the
radiative corrections to $m_{h_{125}}$ mainly
arise. Related to $m_{h_{125}}$, one can
also see that large parts of
the green area, which indicates
a worse fit to the signal-rate measurements
of $h_{125}$, overlap with the
pink area. The green area
is defined as the parameter region in which the
points have $\Delta \chi^2=
\chi^2_{\mu\nu{\rm SSM}} - \chi^2_{\rm SM}
> 5.99$, 
i.e.\ the points are 
disfavored at the $95\%$ confidence
level in the two-dimensional parameter scan
considered here. In contrast to the exclusions
from direct searches, the exclusion limits related
to the properties of $h_{125}$ (pink and
green areas) stretch out over 
parts of the
blue parameter region.

\section{Conclusions and outlook}
\label{conclu}
In this paper we have presented an
investigation of the stability of
the EW vacuum of the $\mu\nu$SSM.
We have described in detail the
approach that was used in order
to determine the dangerous unphysical
minima in the eight-dimensional
field space of the CP-even scalar
fields. Moreover, we have utilized
a numerically robust semi-classical
approximation for the computation
of the EW vacuum decay rates, which allowed us
to categorize
parameter points for which the
EW minimum is not the global
minimum into points featuring a
metastable long-lived or an unstable
EW vacuum (as well as an intermediate
region between the two cases).

Focusing on the alignment without
decoupling limit of the model,
we have demonstrated in our numerical
discussion how the analysis of
vacuum stability can provide important
constraints on the parameter space.
We have generically found that the
EW minimum is not the global minimum
of the potential. In comparison to
the NMSSM, 
the
presence of three gauge singlet
scalar fields in the $\mu\nu$SSM leads to
the possibility of 
various potentially
dangerous unphysical minima existing
in the neutral scalar potential
for a single parameter point.
We have found
that the largest rates
of the EW vacuum decays are related
to the unphysical
minima that are the closest in field
space to the EW minimum, despite the
fact that there might be even
deeper unphysical minima present.
Furthermore, our results clearly indicate that
simply requiring that the EW minimum is
the global minimum of the potential would
exclude large parts of the
$\mu\nu$SSM parameter space that
are actually phenomenologically viable. 
Instead, it is crucial to take into account
the possibility of a metastable EW vacuum 
and to perform an analysis of
the EW vacuum decay rates.

We have shown analytically and
numerically that an
unstable EW vacuum can be expected to
occur quite generically in regions of the
parameter space in which relatively
light right-handed sneutrinos are
present. Due to the fact that these
particles are gauge singlets, thus
not coupled directly to the SM particles,
they 
may not be detectable
at
collider experiments even if they
have small masses. It is therefore
an important finding that the corresponding
parameter space can be constrained
in other
ways, such as the
vacuum stability investigation as
presented here.
On the other hand, 
for the case where 
the
sneutrinos 
have a significant coupling
to the
SM particles via a mixing with the
Higgs boson at $125\gev$, constraints
from collider experiments in
combination with constraints from the
analysis of the EW vacuum stability can
be utilized in a complementary way
in order to exclude parts of the
parameter space of the $\mu\nu$SSM.

As an outlook to possible
future
research, we emphasize that 
further constraints on the parameter space of the $\mu\nu$SSM 
may be obtained by complementing the present analysis at $T = 0$ with 
an investigation of the thermal history of the universe, 
see recent studies
in other
scalar extensions of the
SM~\cite{Baum:2020vfl,Biekotter:2021ysx}.
While the
EW minimum might be sufficiently long-lived
at zero temperature,
the universe might
have adopted one of the unphysical
minima 
during its thermal history.
It can then happen that the transition rate
for the phase transition into the
EW vacuum 
would have never been
large enough to
allow for the onset of EW symmetry breaking,
and the universe would have been trapped until
zero temperature in one of the unphysical
minima. This \textit{vacuum
trapping} effect is especially
important if the formation of vacua
at the origin, or vacua
in which only the singlet fields obtain
vacuum expectation values, are possible,
and we encountered such cases in our analysis.
It should be noted
that minima of the latter kind usually
form at temperatures much larger than
the EW scale, because vanishing field
values of the singlet fields are not stabilized
via the SM interactions.
In the event that
an unphysical minimum at the origin
(or any minimum 
that features
non-zero vevs only for the right-handed
sneutrino fields)
can exist until zero temperature there are then
two possibilities: The universe is trapped
in such a vacuum, or the universe
undergoes a first-order EW phase transition.
While the vacuum trapping might
yield important 
constraints on the
parameter space of the $\mu\nu$SSM, the
possibility of first-order EW phase transitions
can facilitate baryogenesis or give rise to the
formation of a stochastic gravitational wave
background in the early universe,
thus pointing towards
particularly interesting regions of
parameter space.
An investigation
of the thermal history of the $\mu\nu$SSM
is left for future work.\footnote{See
\citere{Chung:2010cd} for a discussion of
the possibility of
first-order EW phase transitions in the
$\mu\nu$SSM.}

\section*{Acknowledgements}
We thank F.~Campello,
D.~L\'opez-Fogliani,
C.~Mu\~noz and
J.~Wittbrodt for interesting discussions.
The work of T.B.\ and G.W.\ is supported by the Deutsche
Forschungsgemeinschaft under Germany’s Excellence
Strategy EXC2121 ``Quantum Universe'' - 390833306.
The work of S.H.\ is supported in part by the
MEINCOP Spain under contract PID2019-110058GB-C21
and in part by
the AEI through the grant IFT Centro de
Excelencia Severo Ochoa SEV-2016-0597.

\appendix

\section{Checking vacuum stability with
\texttt{\textbf{munuSSM}}}
\label{appcode}
The determination of the unphysical minima and the
computation of the
lifetime of (metastable) EW vacuum
in the $\mu\nu$SSM can be
performed with the new version of the
public code \texttt{munuSSM}. The
implementation follows the approach
described in \refse{vacstab}, which is
based on \citere{Hollik:2018wrr}.
The stationary
conditions are solved with an interface to
the public code
\texttt{Hom4PS2}~\cite{Lee:hom4ps2}.\footnote{The
code can be downloaded precompiled
at \url{http://www.math.nsysu.edu.tw/~leetsung/works/HOM4PS_soft.htm}.}
We briefly summarize here the usage of the
code employed for the analysis
presented in \refse{numana}, including
a description of the user interface
of the new vacuum stability test.
A more detailed account of
the user interface can be found
in the documentation under
\url{https://www.desy.de/~biek/munussmdocu/site/}.

Assuming that there is an input file
called \texttt{path} in which the values
for the free parameters are defined
(see \citere{Biekotter:2020ehh} for details),
the corresponding
parameter point can be analyzed by executing
the commands:
\begin{lstlisting}
from munuSSM.benchmarkPointFromFile import \
    BenchmarkPointFromFile
from munuSSM.vacuumStability.checkPotential import \
    CheckPotential
from munuSSM.higgsBounds.util import \
    check_higgsbounds_higgssignals

pt = BenchmarkPointFromFile(file='path')
pt.calc_loop_masses()
checker = CheckPotential(pt)
checker.check_stability()
check_higgsbounds_higgssignals(pt)
\end{lstlisting}
The results of the \texttt{HiggsBounds}
and the \texttt{HiggsSignals} test are then
saved in the dictonaries:
\begin{lstlisting}
pt.HiggsBounds
pt.HissSignals
\end{lstlisting}
The definition of each item of the
dictionaries can be found in
\citere{Biekotter:2020ehh}.
The most time-consuming call is the
one of the function \texttt{check\_stability},
which depending on the input parameters
usually takes roughly a minute to finish.
The results of the vacuum stability
check are saved in:
\begin{lstlisting}
pt.local_minima
pt.Transitions
\end{lstlisting}
The attribute \texttt{local\_minima} is a list
of dictionaries, where each dictionary
saves the field values $\phi_d$
and the potential
depth $V(\phi_d)$ of each local minimum.
The attribute \texttt{Transitions} is
also a list of dictionaries, where each
dictionary saves the relevant information
about the possible transitions into each
deeper unphysical minimum $\phi_d$.
These dictionaries have the following items:
\vspace*{-0.6em}
\begin{multicols}{3}
\begin{itemize}[noitemsep,topsep=0pt]
\item[] \texttt{'DangMin'}:
    $\phi_d$
\item[] \texttt{'V'}:
    $V(\phi_d)$
\item[] \texttt{'unitvec'}:
    $\hat \varphi_d$
\item[] \texttt{'unit\_min\_pos'}:
    $\Delta \phi_d$
\item[] \texttt{'msq'}:
    $m^2(\hat \varphi_d)$
\item[] \texttt{'A'}:
    $A(\hat \varphi_d)$
\item[] \texttt{'lambda'}:
    $\lambda(\hat \varphi_d)$
\item[] \texttt{'delta'}:
    $\delta(\hat \varphi_d)$
\item[] \texttt{'Action'}:
    $S_E$
\end{itemize}
\end{multicols}
\vspace*{-0.6em}
\noindent As discussed in \refse{vacstab},
a parameter point can be considered
to be valid if all values of $S_E$
are larger than $440$.
If the attribute \texttt{Transitions}
is empty, then there are no
unphysical minima below the EW minimum,
and the point is also valid.

\section{Coefficients of 
\texorpdfstring{\boldmath{$V_{\rm EW}$}}{VEW} for special cases} 
\label{app:coefs}
We consider here two special cases, for which we provide 
explicit expressions for the coefficients of the scalar potential 
relative to the EW vacuum, $V_{\rm EW}(\varphi, \hat\varphi)$, 
as given in \refeq{eq1dpot}. 
If one of the right-handed sneutrino fields
$\phi_{iR}$
is mixed with the SM-like Higgs boson, then
transitions along the directions $-\phi_{iR}$
and $\phi_u$ occur
(see AII, BIII and CIV in \refta{tab:unphy}).
Choosing for instance $i = 1$,
the unit vector $\hat \varphi$
for such transitions can be
approximately expressed as
\begin{equation}
\hat\varphi =
N
\left(0,
D,
1 / D,
0,
0,
0, 0, 0 \right) \ , \; 
\text{with} \;
N = \frac{D}{\sqrt{D^4 + 1}} \ ,
\end{equation}
where $D$ parametrizes the 
direction in field space, and
the variation of the other fields
was neglected.
Then the coefficients of
$V_{\rm EW}$ are given by
\begin{dmath}
m^2 = -\frac{\lambda _1 \lambda _2 N^2 v^2 c_{\beta }^2 v_{2 R}}{4 D^2 v_{1 R}}-\frac{\lambda _1 \lambda _2
   N^2 v^2 s_{\beta }^2 v_{2 R}}{4 D^2 v_{1 R}}-\frac{\lambda _1 \lambda _3 N^2 v^2 c_{\beta }^2 v_{3
   R}}{4 D^2 v_{1 R}}-\frac{\lambda _1 \lambda _3 N^2 v^2 s_{\beta }^2 v_{3 R}}{4 D^2 v_{1
   R}}+\frac{\lambda _1 N^2 v^2 A_1^{\lambda } c_{\beta } s_{\beta }}{2 \sqrt{2} D^2 v_{1
   R}}+\frac{D^2 \lambda _1 N^2 A_1^{\lambda } c_{\beta } v_{1 R}}{2 \sqrt{2} s_{\beta
   }}+\frac{\kappa _1 N^2 A_1^{\kappa } v_{1 R}}{2 \sqrt{2} D^2}+\frac{D^2 \kappa _1 \lambda _1 N^2
   c_{\beta } v_{1 R}^2}{4 s_{\beta }}+\kappa _1 \lambda _1 N^2 v c_{\beta } v_{1 R}+\frac{\kappa
   _1^2 N^2 v_{1 R}^2}{D^2}-\lambda _1^2 N^2 v s_{\beta } v_{1 R}+\frac{D^2 \lambda _2 N^2
   A_2^{\lambda } c_{\beta } v_{2 R}}{2 \sqrt{2} s_{\beta }}+\frac{D^2 \kappa _2 \lambda _2 N^2
   c_{\beta } v_{2 R}^2}{4 s_{\beta }}-\lambda _1 \lambda _2 N^2 v s_{\beta } v_{2 R}+\frac{D^2
   \lambda _3 N^2 A_3^{\lambda } c_{\beta } v_{3 R}}{2 \sqrt{2} s_{\beta }}+\frac{D^2 \kappa _3
   \lambda _3 N^2 c_{\beta } v_{3 R}^2}{4 s_{\beta }}-\lambda _1 \lambda _3 N^2 v s_{\beta } v_{3
   R}+\frac{\lambda _1 N^2 v A_1^{\lambda } c_{\beta }}{\sqrt{2}}+\frac{1}{8} D^2 g_1^2 N^2 v^2
   s_{\beta }^2+\frac{1}{8} D^2 g_2^2 N^2 v^2 s_{\beta }^2
   \ ,
\end{dmath}
\begin{dmath}
A =
\frac{\kappa _1^2 N^3 v_{1 R}}{D^3}+\frac{\kappa _1 N^3 A_1^{\kappa }}{3 \sqrt{2} D^3}+\frac{\kappa
   _1 \lambda _1 N^3 v c_{\beta }}{2 D}+\frac{1}{2} D \lambda _1 \mu  N^3-\frac{\lambda _1^2 N^3 v
   s_{\beta }}{2 D}
   \ ,
\end{dmath}
\begin{dmath}
\lambda =
\frac{1}{32} D^4 g_1^2 N^4+\frac{1}{32} D^4 g_2^2 N^4+\frac{\kappa _1^2 N^4}{4 D^4}+\frac{1}{4}
   \lambda _1^2 N^4 \ .
\end{dmath}
Another kind of transition that was
encountered in the numerical discussion
was one in which two of the right-handed
sneutrino fields, e.g., $\phi_{1R}$ and
$\phi_{2R}$, evolve from $v_R$ to
approximately zero (see BI and CI 
in \refta{tab:unphy}).
Neglecting the variation of the Higgs
doublet fields $\phi_d$ and $\phi_u$
for these transitions, one can parametrize
the unit vector by
\begin{equation}
\hat\varphi =
-N
\left( 0, 0,
D, 1 / D,
0, 0, 0, 0 \right) \ ,
\end{equation}
and one finds for the coefficients
\begin{dmath}
m^2 =
-\frac{D^2 \lambda _1 \lambda _2 N^2 v^2 c_{\beta }^2 v_{2 R}}{4 v_{1 R}}-\frac{\lambda _1 \lambda _2
   N^2 v^2 c_{\beta }^2 v_{1 R}}{4 D^2 v_{2 R}}-\frac{D^2 \lambda _1 \lambda _2 N^2 v^2 s_{\beta }^2
   v_{2 R}}{4 v_{1 R}}-\frac{\lambda _1 \lambda _2 N^2 v^2 s_{\beta }^2 v_{1 R}}{4 D^2 v_{2
   R}}-\frac{D^2 \lambda _1 \lambda _3 N^2 v^2 c_{\beta }^2 v_{3 R}}{4 v_{1 R}}-\frac{D^2 \lambda _1
   \lambda _3 N^2 v^2 s_{\beta }^2 v_{3 R}}{4 v_{1 R}}+\frac{D^2 \lambda _1 N^2 v^2 A_1^{\lambda }
   c_{\beta } s_{\beta }}{2 \sqrt{2} v_{1 R}}+\frac{D^2 \kappa _1 N^2 A_1^{\kappa } v_{1 R}}{2
   \sqrt{2}}+D^2 \kappa _1^2 N^2 v_{1 R}^2-\frac{\lambda _2 \lambda _3 N^2 v^2 c_{\beta }^2 v_{3
   R}}{4 D^2 v_{2 R}}-\frac{\lambda _2 \lambda _3 N^2 v^2 s_{\beta }^2 v_{3 R}}{4 D^2 v_{2
   R}}+\frac{\lambda _2 N^2 v^2 A_2^{\lambda } c_{\beta } s_{\beta }}{2 \sqrt{2} D^2 v_{2
   R}}+\frac{\kappa _2 N^2 A_2^{\kappa } v_{2 R}}{2 \sqrt{2} D^2}+\frac{\kappa _2^2 N^2 v_{2
   R}^2}{D^2}+\frac{1}{2} \lambda _1 \lambda _2 N^2 v^2 c_{\beta }^2+\frac{1}{2} \lambda _1 \lambda
   _2 N^2 v^2 s_{\beta }^2 \ ,
\end{dmath}
\begin{dmath}
A =
D^3 \kappa _1^2 N^3 v_{1 R}+\frac{\kappa _2^2 N^3 v_{2 R}}{D^3}+\frac{D^3 \kappa _1 N^3 A_1^{\kappa
   }}{3 \sqrt{2}}+\frac{\kappa _2 N^3 A_2^{\kappa }}{3 \sqrt{2} D^3}+\frac{D \kappa _1 N^3
   A_1^{\kappa }}{\sqrt{2}}+\frac{\kappa _1 N^3 A_1^{\kappa }}{\sqrt{2} D} \ ,
\end{dmath}
\begin{dmath}
\lambda =
\frac{1}{4} D^4 \kappa _1^2 N^4+\frac{\kappa _2^2 N^4}{4 D^4}
\ .
\end{dmath}
As a cross check one can 
use the feature
that since both $\phi_{1R}$ and
$\phi_{2R}$ are gauge-singlet fields,
the expressions shown above are invariant
under the transformation $D \to 1/D$
and at the same time exchanging
the indices $1 \leftrightarrow 2$.

\bibliographystyle{utphys}
\bibliography{lit}

\end{document}